\documentclass{PoS-hep}
\usepackage{amsmath}
\usepackage{bm}
\usepackage{epsfig}
\usepackage{graphics}
\usepackage{upgreek}
\usepackage{amsfonts}
\usepackage{amsbsy}
\usepackage{amscd}
\usepackage{bbm}
\usepackage{eufrak}
\usepackage{cite}

\renewenvironment{subequations}{%
\refstepcounter{equation}%
\setcounter{parentequation}{\value{equation}}%
  \setcounter{equation}{0}
  \ignorespaces
}{%
  \setcounter{equation}{\value{parentequation}}%
  \ignorespacesafterend
}
\newcommand{\eeq}{\end{equation}}
\newcommand{\beq}{\begin{equation}}
\newcommand{\ba}{\begin{array}}
\newcommand{\ea}{\end{array}}
\newcommand{\bea}{\begin{eqnarray}}
\newcommand{\eea}{\end{eqnarray}}

\newcommand{\beqs}{\begin{subequations}}
\newcommand{\eeqs}{\end{subequations}}
\newcommand{\eec}{\end{center}}
\newcommand{\bec}{\begin{center}}
\newcommand{\eem}{\end{matrix}}
\newcommand{\bem}{\begin{matrix}}
\newcommand{\Eref}[1]{Eq.~(\ref{#1})}
\newcommand{\Sref}[1]{Sec.~\ref{#1}}
\newcommand{\Fref}[1]{Fig.~\ref{#1}}
\newcommand{\Tref}[1]{Table~\ref{#1}}
\newcommand{\cref}[1]{Ref.~\cite{#1}}

\newcommand{\sEref}[2]{Eq.~(\ref{#1}{\ftn\sf {#2}})}

\newcommand{\ftn}{\footnotesize}

\newcommand{\TeV}{{\mbox{\rm TeV}}}

\newcommand{\GeV}{{\mbox{\rm GeV}}}

\newcommand{\pb}{{\mbox{\rm pb}}}

\def\lf{\left(}
\def\rg{\right)}

\newcommand{\hh}{{\ensuremath{I{\kern-2.6pt h}}}}
\newcommand{\bhh}{{\ensuremath{\bar{I{\kern-2.6pt h}}}}}

\newcommand{\gev}{~\text{GeV}}
\newcommand{\tev}{~\text{TeV}}

\newcommand{\omg}{\Omega_{\mathrm{LSP}}h^2}

\newcommand{\damu}{\delta a_\mu}
\newcommand{\dstau}{\Delta_{\tilde{\tau}_1}}
\newcommand{\dch}{\Delta_{\tilde{\chi}^+_1}}
\newcommand{\dnt}{\Delta_{\tilde{\chi}_2}}
\newcommand{\ssi}{\sigma_{\tilde{\chi}p}^{\mathrm{SI}}}
\newcommand{\xssi}{\xi\sigma_{\tilde{\chi}p}^{\mathrm{SI}}}
\newcommand{\ssd}{\sigma_{\tilde{\chi}p}^{\mathrm{SD}}}
\newcommand{\amg}{A_0/M_{1/2}}

\newcommand{\DR}{$\overline{\mathrm{DR}}~$}
\newcommand{\MS}{$\overline{\mathrm{MS}}~$}
\newcommand{\neutralino}{\tilde{\chi}}

\newcommand{\staua}{\tilde{\tau}_1}

\newcommand{\bsmumu}{B_s\rightarrow\mu^+\mu^-}
\newcommand{\bsmm}{{\ensuremath{{\rm BR}\lf B_s\to \mu^+\mu^-\rg}}}
\newcommand{\bsg}{{\ensuremath{{\rm BR}\lf b\to s\gamma\rg}}}
\newcommand{\btn}{{\ensuremath{{\rm R}\lf B_u\to \tau\nu\rg}}}
\newcommand{\Omx}{{\ensuremath{\Omega_{\rm LSP} h^2}}}
\newcommand{\mx}{{\ensuremath{m_{\rm LSP}}}}
\newcommand{\mgl}{{\ensuremath{m_{\tilde g}}}}
\newcommand{\mch}{{\ensuremath{m_{\tilde \chi^\pm}}}}

\newcommand{\cha}{{\ensuremath{{\tilde \chi^+_1}}}}
\newcommand{\mh}{{\ensuremath{m_{h}}}}
\newcommand{\mo}{{\ensuremath{m_{0}}}}
\newcommand{\mgo}{\ensuremath{\mg>\mo}}
\newcommand{\mog}{\ensuremath{\mo\gg\mg}}

\newcommand{\Dst}{{\ensuremath{\Delta_{\tilde\tau_1}}}}
\newcommand{\dew}{{\ensuremath{\Delta_{\rm EW}}}}

\newcommand{\mg}{{\ensuremath{M_{1/2}}}}
\newcommand{\xx}{{\ensuremath{\tilde\chi}}}
\newcommand{\xxb}{{\ensuremath{\tilde\chi_2}}}
\newcommand{\xstau}{{\ensuremath{\tilde\chi-\tilde\tau_1}}}
\newcommand{\tnb}{{\ensuremath{\tan\beta}}}
\newcommand{\sign}{{\ensuremath{\rm sign}}}

\title{CMSSM With Generalized Yukawa
Quasi-Unification: An Update}

\ShortTitle{CMSSM with Generalized Yukawa Quasi-Unification: An
Update}

\author{N. Karagiannakis\\ School of Electrical and Computer
Engineering, Faculty of Engineering, \\ Aristotle University of
Thessaloniki, Thessaloniki 54124, GREECE \\ E-mail:
\email{nikar@auth.gr}}

\author{G. Lazarides\\ School of Electrical and Computer Engineering, Faculty
of Engineering, \\ Aristotle University of Thessaloniki,
Thessaloniki 54124, GREECE \\ E-mail:
\email{lazaride@eng.auth.gr}}

\author{\speaker{C. Pallis}\\
Departament de F\'isica Te\`orica and IFIC,\\
Universitat de Val\`encia-CSIC, E-46100 Burjassot, SPAIN\\
E-mail: \email{cpallis@ific.uv.es}}

\abstract{We analyze the parametric space of the constrained
minimal supersymmetric standard model (CMSSM) with $\mu>0$
supplemented by a generalized asymptotic Yukawa coupling
quasi-unification condition which yields acceptable masses for the
fermions of the third family. We impose constraints from the cold
dark matter abundance in the universe and its direct detection
experiments, the $B$-physics, as well as the masses of the
sparticles and the lightest neutral CP-even Higgs boson, $\mh$. We
identify two distinct allowed regions with $\mgo$ and $\mog$
classified in the hyperbolic branch of the radiative electroweak
symmetry breaking. In the first region we obtain, approximately,
$44\lesssim\tan\beta\lesssim52$, $-3\lesssim \amg\lesssim0.1$,
$122\lesssim\mh/\GeV\lesssim127$, and mass of the lightest
sparticle in the range $(0.75-1.43)~\TeV$. Such heavy lightest
sparticle masses can become consistent with the cold dark matter
requirement on the lightest sparticle relic density thanks to
neutralino-stau coannihilations. In the latter region, fixing
$\mh$ to its central value from the LHC, we find a wider allowed
parameter space with milder electroweak-symmetry-breaking
fine-tuning, $40\lesssim\tnb\lesssim50$,
$-11\lesssim\amg\lesssim15$ and mass of the lightest sparticle in
the range $(0.09-1.1)~\TeV$. This sparticle is possibly detectable
by the present cold dark matter direct search experiments.
\\\\
{\sl\bfseries Published in}~~{PoS  CORFU {\bf 2014}, 115 (2015)}.
}

\FullConference{Proceedings of the Corfu Summer Institute 2014 \\
                 3-21 September 2014\\
                 Corfu, Greece}

\begin{document}

\section{Introduction}

One of the most economical and predictive versions of the
\textit{minimal supersymmetric standard model} ({\sf\ftn  MSSM})
is the well-known -- see e.g \cref{barger1994} --
\textit{constrained MSSM} ({\sf\ftn CMSSM}) which employs just
four and one half free parameters:
\begin{equation}
\sign\mu,~~\tan\beta,~~\mg,~~m_0,~~\mbox{and}~~A_0, \label{par}
\end{equation}
where $\sign\mu$ is the sign of $\mu$, the mass parameter mixing
the electroweak Higgs superfields $H_2$ and $H_1$ of the MSSM
which couple to the up- and down-type quarks respectively, $\tnb$
is the ratio of the vacuum expectation values of $H_2$ and $H_1$,
while the remaining symbols denote the common gaugino mass, scalar
mass, and trilinear scalar coupling, respectively, defined at the
\emph{grand unified theory} ({\sf\ftn GUT}) scale $M_{\rm GUT}$,
which is determined by the unification of the gauge coupling
constants.

CMSSM can be further restricted by being embedded in a
\emph{supersymmetric} ({\sf\ftn  SUSY}) GUT with a gauge group
containing $SU(4)_{\rm c}$ and $SU(2)_R$. This can lead to
`asymptotic' \emph{Yukawa unification} ({\sf\ftn  YU}) \cite{als},
i.e. the exact unification of the third generation Yukawa coupling
constants at the SUSY GUT scale $M_{\rm GUT}$. Given the
experimental values of the top-quark and tau-lepton masses, the
CMSSM supplemented by the assumption of YU (which naturally
restricts $\tan\beta\sim50$) yields unacceptable values
\cite{copw} of the $b$-quark mass for both signs of the MSSM
parameter $\mu$. In Ref.~\cite{qcdm} -- see also
Refs.~\cite{oxford,nova,ijmp} --, this problem is addressed in the
context of the \emph{Pati-Salam} ({\sf\ftn PS}) GUT model, without
the need of invoking departure from the CMSSM universality. We
prefer to sacrifice the exact YU in favor of the universality
hypothesis, since we consider the latter as more economical,
predictive, and easily accommodated within conventional SUSY GUT
models. In particular, the Higgs sector of the simplest PS model
\cite{antoniadis1989} is extended so that $H_2$ and $H_1$ are not
exclusively contained in a $SU(4)_c$ singlet, $SU(2)_{\rm L}\times
SU(2)_{\rm R}$ bidoublet superfield, but receive subdominant
contributions from another bidoublet too which belongs to the {\bf
15} representation of $SU(4)_c$. As a result, YU is naturally
violated and replaced by a set of asymptotic \emph{Yukawa
quasi-unification conditions} ({\sf\ftn YQUCs}):
\beq\label{yquc}
 h_t(M_\mathrm{GUT}):h_b(M_\mathrm{GUT}):h_\tau(M_\mathrm{GUT})=
\left|\frac{1-\rho\alpha_2/\sqrt{3}}{\sqrt{1+|\alpha_2|^2}}\right|:
\left|\frac{1-\rho\alpha_1/\sqrt{3}}{\sqrt{1+|\alpha_1|^2}}\right|:
\left|\frac{1+\sqrt{3}\rho\alpha_1}{\sqrt{1+|\alpha_1|^2}}\right|.
\eeq
These conditions depend on two complex parameters $\alpha_1$,
$\alpha_2$ and one real and positive parameter $\rho$. The
parameters $\alpha_1$ and $\alpha_2$ describe the mixing of the
components of the $SU(4)_\mathrm{c}$ singlet and 15-plet Higgs
bidoublets, while $\rho$ is the ratio of their respective Yukawa
coupling constants to the fermions of the third family. Note that
monoparametric versions of the YQUCs considered within CMSSM
arising by taking $\alpha_1=-\alpha_2$ for $\mu>0$ \cite{nikos11}
or $\alpha_1=\alpha_2$ for $\mu<0$ \cite{gomez2003} are by now
experimentally excluded \cite{gomez2003,pekino}. In this talk,
based on \cref{nikos13,nikos15}, we show that the YQUCs in
\Eref{yquc} can become compatible with two disconnected regions of
the CMSSM parameter space belonging to the \emph{hyperbolic
branch} ({\sf\ftn HB}) of the radiative \emph{electroweak symmetry
breaking} ({\sf\ftn EWSB}): The $\mgo$ region where the
neutralino, $\xx$, is a pure bino and may (co)annihilate strongly
enough with the lighter stau, $\staua$ \cite{cmssm1,cdm}; and the
$\mog$ region, where $\xx$ acquires a sizable higgsino fraction
\cite{baerfocus, nathfocus,feng, akula2012, nath13} which enhances
the $\xx-\xx$ annihilation and triggers neutralino-chargino
($\xx/\xxb-\cha$) coannihilations.

We begin by describing the cosmo- \& phenomeno- logical
requirements which we consider in our investigation in
Sec.~\ref{sec:pheno}. Next (\Sref{res}), we exhibit the resulting
restrictions on the parameters of the CMSSM. Moreover, we check in
\Sref{yuk} the consistency with \Eref{yquc} and discuss the
naturalness of the model in \Sref{nat}. We summarize our
conclusions in Sec.~\ref{con}.

\section{Phenomenological and Cosmological Constraints}
\label{sec:pheno}

In our investigation, we integrate the two-loop renormalization
group equations for the gauge and Yukawa coupling constants and
the one-loop ones for the soft SUSY breaking parameters between
$M_{\rm GUT}$ and a common SUSY threshold $M_{\rm SUSY}
\simeq(m_{\tilde t_1}m_{\tilde t_2})^{1/2}$ ($\tilde t_{1,2}$ are
the stop mass eigenstates) determined consistently with the SUSY
spectrum. At $M_{\rm SUSY}$, we impose radiative EWSB, evaluate
the SUSY spectrum employing the publicly available calculator {\tt
SOFTSUSY} \cite{allanach2002}, and incorporate the SUSY
corrections to the $b$ and $\tau$ mass -- see below. From $M_{\rm
SUSY}$ to $M_Z$, the running of gauge and Yukawa coupling
constants is continued using the \textit{standard model} ({\sf\ftn
SM}) renormalization group equations. The SUSY spectrum is put
into \texttt{micrOMEGAs} \cite{micro}, a publicly available code
which calculates a number of phenomenological -- see \Sref{pheno}
-- and cosmological -- see \Sref{cosmo} -- observables which
assist us to restrict the parametric space of our model.

\subsection{Phenomenological Requirements} \label{pheno}

\paragraph{2.1.1 SM Fermion Masses.} After incorporating the sizable (about 20\%) and less important
(almost 4$\%$) corrections \cite{pierce1997} to the $b$-quark and
$\tau$-lepton masses, we compare the masses of top-quark, $m_t$,
$b$-quark, $m_b$ and $\tau$-lepton, $m_\tau$ with their
experimental values\cite{particledata,mtnew}
\begin{equation}
m_t(m_t)=164.83\gev,~~
m_b(m_b)^\text{\MS}=4.18\gev,~~m_\tau(M_Z)=1.748\gev\label{pfm}.
\end{equation}
The second value is evolved up to $M_Z$ using the central value
$a_s(M_Z)=0.1185$ \cite{particledata} of the strong fine-structure
constant at $M_Z$ and then converted to the \DR scheme with result
$m_b(M_Z)=2.83\gev$.

\paragraph{2.1.2 Collider Bounds.}

For our analysis, the relevant collider bounds constrain:

\begin{itemize}

\item{The mass $\mh$ of the lightest Higgs boson, $h$.} The
experiments ATLAS \cite{Aad:2014aba} and CMS \cite{CMS:2014ega} in
the LHC discovered simultaneously a boson that looks very much
like the expected SM Higgs boson. The allowed $95\%$
\emph{confidence level} ({\sf\ftn c.l.}) range of $\mh$ can be
estimated including a theoretical uncertainty of about
$\pm1.5\gev$. This gives
\begin{equation}
\label{mhb}  122\lesssim \mh/\GeV\lesssim128.5.
\end{equation}

\item{The masses of the lightest chargino, $\mch$,
\cite{LEPchargino} and gluino, $\mgl$ \cite{ATLAScol2013}:}
\begin{equation}\label{mchb}
  \mbox{\sf\ftn (a)}\>\>\>\mch\gtrsim103.5\gev~~~\mbox{and}~~~\mbox{\sf\ftn (b)}\>\>\>\mgl\gtrsim1.3\tev.
\end{equation}

\end{itemize}

\paragraph{2.1.3 {\sl\large B}-Physics Constraints.} \label{pbph}

SUSY contributions to observables related to $B$-meson physics
yield restrictions to the SUSY parameters. In particular, we
impose the following bounds on:

\begin{itemize}

\item The branching ratio $\bsmm$ of the process
$B_s\to\mu^+\mu^-$ \cite{LHCb2012,albrecht2012}
\begin{equation}\label{bmmb}
 \mathrm{BR}(\bsmumu)\lesssim4.2\times10^{-9}.
\end{equation}

\item The branching ratio $\bsg$ of $b\to s\gamma$ \cite{bsgexp,
bsgSM}:
\begin{equation}
  2.79\times 10^{-4}\lesssim \bsg \lesssim 4.07\times 10^{-4}.
\label{bsgb} \end{equation}

\item The ratio $\btn$ of the CMSSM to the SM branching ratio of
the process $B_u\to \tau\nu$ \cite{bsgexp}
\beq 0.52\lesssim\btn\lesssim2.04\,.\label{btnb} \eeq

\end{itemize}

\paragraph{2.1.4 Muon Anomalous Magnetic Moment.}\label{pam}

There is a $2.9-\sigma$
\cite{hagiwara2011,davier2011,muong2col2006} discrepancy
\begin{equation}\label{damub}
\damu=(24.9\pm8.7)\times10^{-10}~~~\Rightarrow~~~
7.5\times10^{-10}\lesssim\damu\lesssim42.3\times10^{-10}~~~\mbox{at
$95\%$ c.l.}
\end{equation}
between the measured value of the muon anomalous magnetic moment
$a_\mu$ from its SM prediction. This $\delta a_\mu$ can be
attributed to SUSY contributions which have the sign of $\mu$ and
its absolute value decreases as $\mx$ increases. Therefore,
\Eref{damub} hints that the sign of $\mu$ has to be positive.
Moreover, a lower [upper] bound on $\mx$ can be derived from the
upper [lower] bound in \Eref{damub}. As it turns out, only the
upper bound on $\mx$ is relevant here. Taking into account the
aforementioned computational instabilities and the fact that a
discrepancy at the level of about $3-\sigma$ cannot firmly
establish a real deviation from the SM value, we restrict
ourselves to just mentioning at which level \Eref{damub} is
satisfied in the parameter space allowed by all the other
constraints.

\subsection{Cold Dark Matter Considerations} \label{cosmo}

\paragraph{2.2.1 CDM Abundance.} In the context of the CMSSM, $\xx$ can be
the \emph{lightest SUSY particle} ({\sf\ftn  LSP}) and, thus,
naturally arises as a \emph{Cold Dark Matter} ({\sf\ftn CDM})
candidate as long as its relic abundance does not exceed the upper
bound on the CDM abundance deduced from the Planck satellite
\cite{Ade:2013zuv}
\begin{equation}\label{omgb}
  \omg\lesssim0.125.
\end{equation}
Two important mechanisms which assist to achieve $\omg$ consistent
with the limit above within CMSSM are {\sf\ftn (i)} the
coannihilation of $\xx$ with a particle $P$ when a proximity
between the mass $\mx$ of $\xx$ and the mass $m_{P}$ of $P$ is
established; {\sf\ftn (ii)} the $P'$-pole effect which enhances
the $\xx-\xx$ pair annihilation procedure by an $P'$-pole exchange
in the $s$-channel when the mass of $P'$ satisfies the relation
$m_{P'}\simeq 2\mx$. The strength of the coannihilation and the
$P'$-pole effect processes is controlled by the relative mass
splittings. The relevant for our cases mass splittings are defined
as follows
\beq \label{ds} \Delta_P=\frac{(m_{P}-m_{\rm LSP})}{m_{\rm
LSP}}~~\mbox{for}~~P=\tilde\tau_1, \cha, \xxb
~~\mbox{and}~~\Delta_{P'}=\frac{m_{P'}-2m_{\rm LSP}}{2m_{\rm
LSP}}~~\mbox{for}~~P'=H\,,\eeq
where $\staua$ is the lightest stau,  $\xxb$ the next-to-lightest
neutralino, $\cha$ the lightest chargino and $H$ the heavier
CP-even neutral Higgs boson. The resulting $\omg$ normally
decreases with these $\Delta_P$'s.

\paragraph{2.2.2 CDM Direct Detection.}

Employing the relevant routine of the {\tt micrOMEGAs} package
\cite{Detmicro} we calculate the \emph{spin-independent} (SI) and
\emph{spin-dependent} (SD) lightest neutralino-proton ($\xx-p$)
scattering cross sections $\ssi$ and $\ssd$, respectively. The
relevant scalar, $f^p_{{{\rm T}_q}}$, and axial-vector,
$\Delta_q^{p}$, form factors for light quarks in the proton (with
$q=u,d,s$), are taken as follows \cite{Detmicro12,latticef}:
\beqs\bea  && f_{{\rm T}u}^p=0.018,~~ f_{{\rm
T}d}^p=0.026,~~\mbox{and}~~f_{{\rm T}s}^p=0.022; \\
\label{Dqp}&& \Delta_{u}^{p}=+0.842,\>\>
\Delta_{d}^{p}=-0.427,\>\>\mbox{and}\>\> \Delta_{s}^{p}=-0.085.
\eea\eeqs
Data on $\ssi$ coming from \textit{large underground Xenon}
({\sf\ftn LUX}) experiment \cite{LUX2014} provide strict bounds on
the values of the free parameters of SUSY models with $\xx$ owning
a sizable higgsino component, as in our case with $\mog$. These
data \cite{brown}, however, are directly applicable in the case
where the CDM consists of just $\xx$'s. If the $\xx$'s constitute
only a part of the CDM in the universe, the LUX experiment bound
on the number of the scattering events is translated into a bound
on the ``rescaled'' SI $\tilde\chi-p$ elastic cross section
$\xi\ssi$, where $\xi= \Omega_{\tilde{\chi}}/0.12$ with $0.12$
being the central value of the CDM abundance \cite{Ade:2013zuv}.

\section{Restrictions on the SUSY Parameters} \label{res}

Imposing the requirements described in \Sref{sec:pheno}, we can
restrict the parameters of our model. Following our approach in
Refs.~\cite{nikos13,nikos15}, we consider as free parameters the
ones in \Eref{par}. The ratios $h_t/h_\tau$ and $h_b/h_\tau$ are
then fixed by using the data of \Eref{pfm}. These ratios must
satisfy the YQUCs in \Eref{yquc} for natural values of the
parameters $\alpha_1,\alpha_2$, and $\rho$ -- see \Sref{yuk}. To
assure this, we restrict ourselves to ratios $h_m/h_n$
($m,n=t,b,\tau$) close to unity which favor the range
$\tnb\geq40$. We also concentrate on the $\mu>0$ case, given that
$\mu<0$ worsens the violation of \Eref{damub}, and scan the region
$-30\leq \amg\leq30$.

We localize below two separated clusters of allowed parameters
categorized in the HB of the radiative EWSB, as justified in
\Sref{res1}: {\ftn\sf (i)} The $\mgo$ area at high $\tnb$
($43.8\lesssim\tnb\lesssim52$) discussed in \Sref{res2} and
{\ftn\sf (ii)} the $\mog$ area for any $\tnb$ in the range $40-50$
studied in \Sref{res3}. We finally -- see \Sref{res4} -- exhibit a
direct comparison of the solutions obtained in the aforementioned
areas focusing on the characteristic value $\tan\beta=48$, which
balances well enough between maintaining natural values for the
$h_m/h_n$'s and satisfying the various requirements of
\Sref{sec:pheno}. Note that the numerical calculations for the
soft SUSY masses become quite unstable for $\mog$ and
$\tan\beta\gtrsim50$.

The constraints which play an important role in delineating both
allowed parameter spaces of our model are the lower bound on $m_h$
in Eq.~(\ref{mhb}) and the CDM bound in Eq.~(\ref{omgb}). The
$\mgo$ region is further restricted by the bound on $\bsmm$ in
Eq.~(\ref{bmmb}) whereas the $\mog$ area is additionally bounded
by the LUX data and the limits on $\mch$ and $\mgl$ in
\sEref{mchb}{a} and {\sf\ftn (b)}. In the parameter space allowed
by these requirements, all the other restrictions of
Sec.~\ref{sec:pheno} are automatically satisfied -- with the
exception of the lower bound on $\delta a_\mu$ in
Eq.~(\ref{damub}).

\subsection{Elliptic Versus Hyperbolic Branch}
\label{res1}

The classification -- see e.g. \cite{akula2012,nath13} -- of the
various solutions of the radiative EWSB condition is based on the
expansion of $\mu^2$ in terms of the soft SUSY breaking parameters
of the CMSSM included in \Eref{par}. Indeed, using fitting
techniques, we can verify the following formula
\beq\mu^2+M_Z^2/2\simeq c_0
m^2_0+c_{1/2}M^2_{1/2}+c_{A}A_0^2+c_{AM}A_0\mg, \label{mufit}\eeq
where the coefficients $c_0, c_{1/2}, c_A$, and $c_{AM}$ depend
basically on $\tnb$ and the masses of the fermions of the third
generation. These coefficients are computed at the scale $M_{\rm
SUSY}$ -- see \Sref{sec:pheno} and, therefore, inherit a mild
dependence on the SUSY spectrum too. From \Eref{mufit}, we can
easily infer that the SUSY breaking parameters are bounded above
for fixed $\mu$, when the quadratic form in the right-hand side of
this equation is positive definite. This is the so-called
\textit{ellipsoidal branch} (EB) which is highly depleted
\cite{nath13} after the discovery of $h$ with $\mh$ in the range
of \Eref{mhb}. On the other hand, in the HB region favored by
\Eref{mhb}, $c_0$ is negative and, consequently, $m_0$ can become
very large together with a combination of $A_0$ and $\mg$ with all
the other parameters being fixed. In this case the soft parameters
lie on \emph{focal} curves or surfaces. Near the boundary between
the EB and HB regions, $c_0$ is very close to zero and, thus,
$m_0$ can become very large with all the other parameters fixed.
These are the so-called \emph{focal points}. Moreover, there is a
region where the soft SUSY breaking mass-squared $m_{H_2}^2$ of
$H_2$ becomes independent of the asymptotic value of the parameter
$m_0$. This is called the \textit{focus point} ({\sf\ftn FP})
region \cite{feng}. In the large $\tnb$ regime under
consideration, we have \cite{barger1994}
$m_{H_2}^2\simeq-\mu^2-M_Z^2/2$ and so no distinction between
focal and focus points can be established.

To get an idea of how our solutions presented in
Secs.~\ref{res2}--\ref{res4} are classified into these categories,
we display in \Tref{tab1} the values of the coefficients $c$ in
\Eref{mufit} for the two representative cases of \Tref{tab2}
corresponding to $\tnb=48$ -- see \Sref{res4}. We obtain $c_0<0$
in both cases and so we expect that our solutions belong to the HB
region. To illustrate the emergence of the relevant focal curves
we diagonalize the quadratic form in the right-hand side of
\Eref{mufit} keeping, e.g., $\mg$ fixed and using $\mo$ and $A_0$
as varying parameters. Then, \Eref{mufit} can be cast in the
following form
\beq m^2_0/C_0+\bar A_0^2/C_A=1 \label{1mufit}\eeq
where $\bar A_0=A+c_{AM}\mg/2c_A$, $C_0=\bar\mu^2/c_0$, and
$C_A=\bar\mu^2/c_A$ with $\bar \mu^2=\mu^2+M_Z^2/2-\bar
c_{1/2}M^2_{1/2}$ and $\bar c_{1/2}=c_{1/2}-c_{AM}^2/4c_A$. The
numerical values of $C_0$ and $C_A$ also listed in \Tref{tab1}.
Since in both cases, one from $C_0$ and $C_A$ is negative, $\mo$
and $A_0$ can vary along an hyperbola parameterized by
\Eref{1mufit} and so both  ($\mgo$ and $\mog$) our solutions in
\Tref{tab2} belong to the HB region.

\begin{table}[t]
\begin{center}
\begin{tabular}{|c|cccc|cc|}\hline
{\sc Region} &  $c_0$
&$c_{1/2}$&$c_{A}$&$c_{AM}$&$C_0/10^8$&$C_A/10^7$\\\hline
$\mgo$& $-0.059$& $0.955$& $0.102$ &$-0.277$&$-1.05$&$6.14$ \\
$\mog$ &  $-0.0689$& $0.848$& $0.0982$ &$-0.254$&$33.3$&$-2.34$
\\\hline
\end{tabular}
\end{center}
\caption{\sl\small The $c$'s in Eq.~(3.1) and $C$'s in Eq.~(3.2)
for $\tnb=48$ and $\mgo$ or $\mog$.}\label{tab1}
\end{table}

\subsection{$\mgo$ Region}
\label{res2}

\begin{figure}[!t]\vspace*{-.12in}
\hspace*{-.19in}
\begin{minipage}{8in}
\epsfig{file=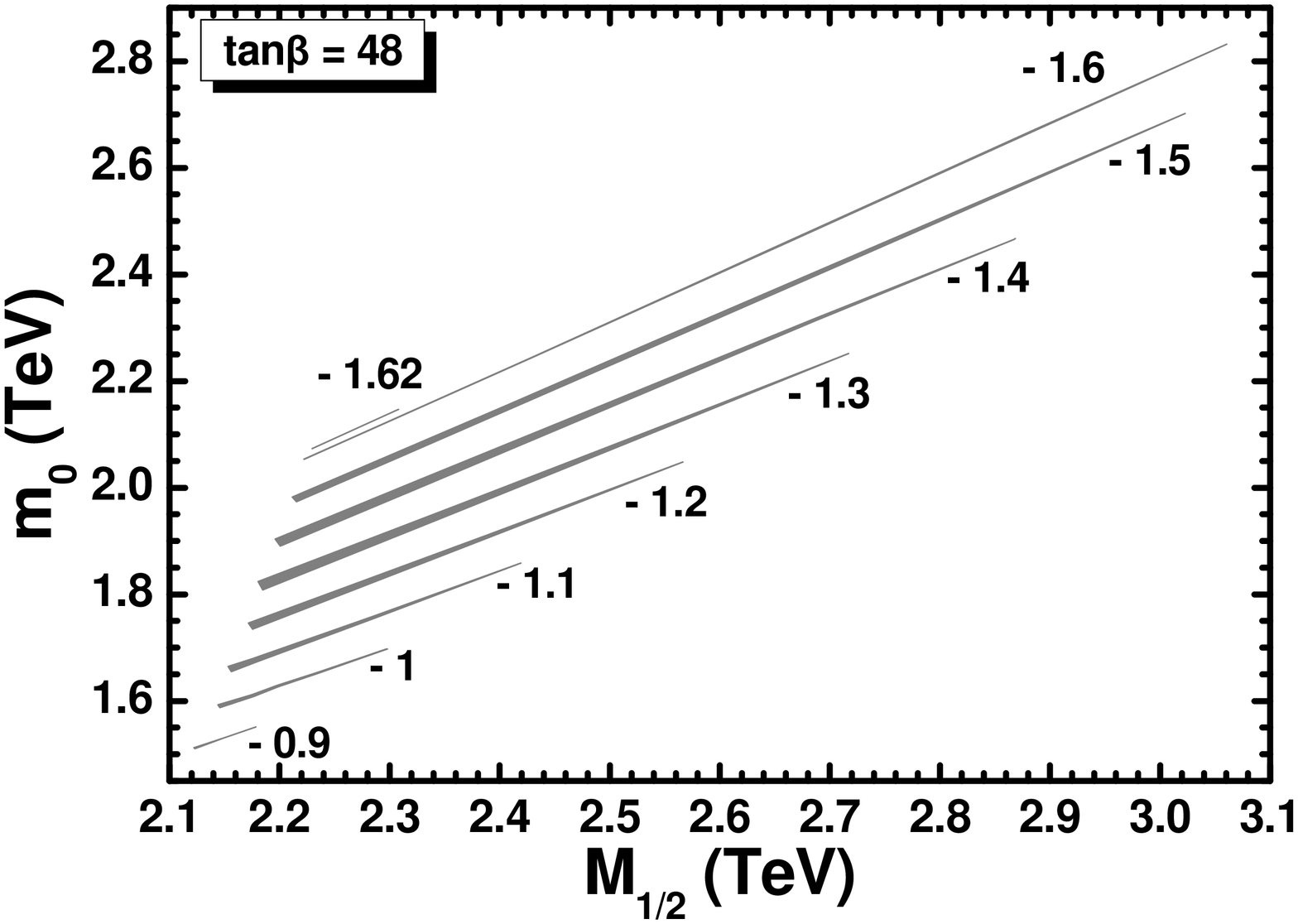,height=3.6in,angle=-90}
\hspace*{-1.2cm}
\epsfig{file=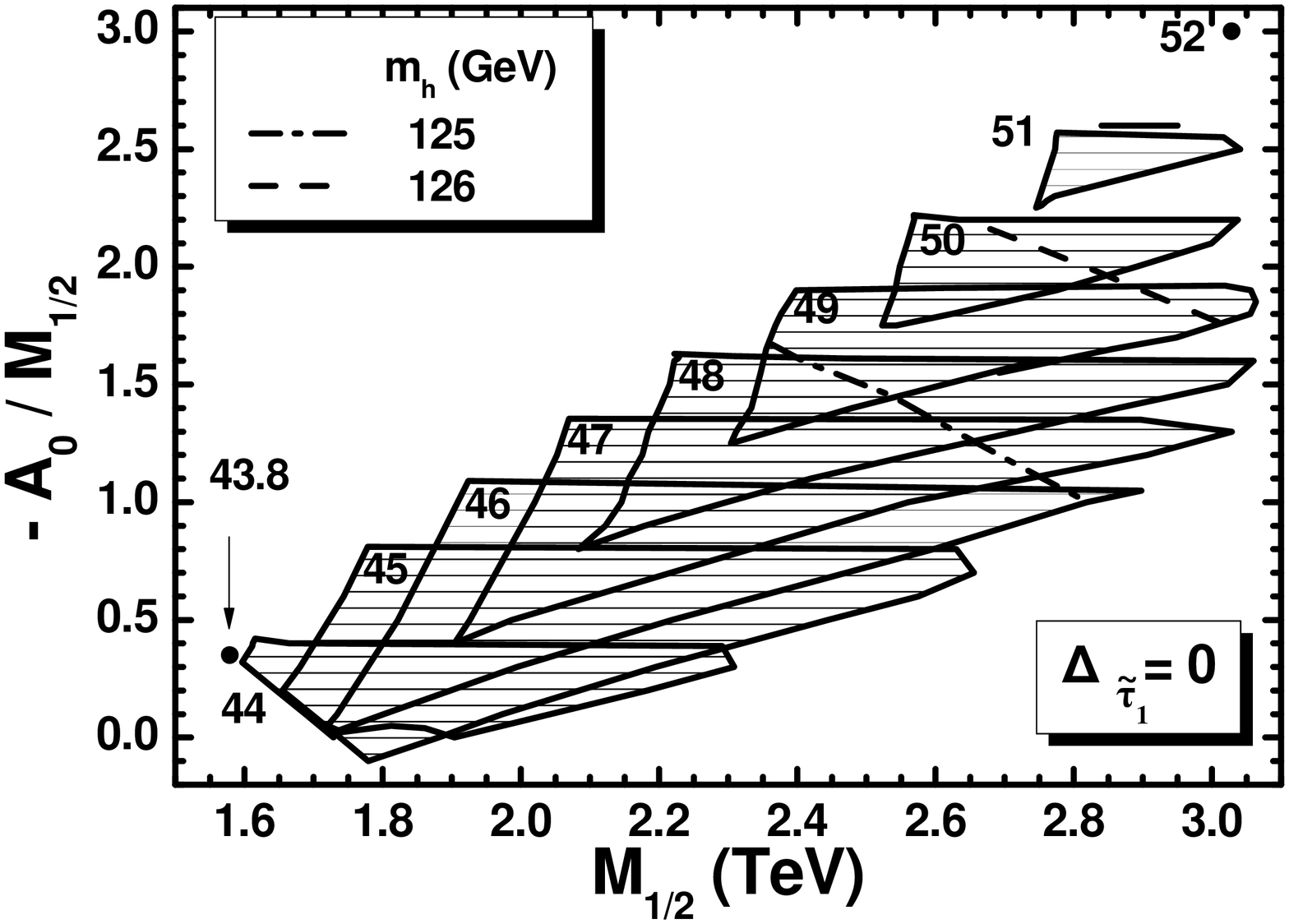,height=3.6in,angle=-90} \hfill
\end{minipage}
\hfill \caption{\sl\small Allowed (shaded) areas in the $\mg-m_0$
[$\mg-\amg$] plane (left [right] plot) for $\mgo$. In the left
[right] plot we use $\tan\beta=48$ [$\Delta_{\tilde\tau_1}=0$] and
various $\amg$'s [$\tan\beta$'s] indicated in the graph. The
dot-dashed [dashed] line corresponds to $m_h=125$ $[126]~{\rm
GeV}$.}\label{fig:tanb48m12m0}
\end{figure}

Initially, we concentrate on the $\mgo$ region and delineate in
the left plot of Fig.~\ref{fig:tanb48m12m0} the allowed (shaded)
areas in the $M_{1/2}-m_0$ plane for $\tan\beta=48$ and various
$\amg$'s indicated therein. The lower boundaries of these areas
corresponds to $\Dst=0$; the areas below these boundaries are
excluded because the LSP is the charged $\tilde\tau_1$. The upper
boundaries of the areas come from the CDM bound in
Eq.~(\ref{omgb}), while the left one originates from the limit on
$\bsmm$ in Eq.~(\ref{bmmb}). The upper right corners of the areas
coincide with the intersections of the lines $\Dst=0$ and
$\Omx=0.125$. We observe that the allowed area, starting from
being just a point at a value of $\amg$ slightly bigger than
$-0.9$, gradually expands as $\amg$ decreases and reaches its
maximal size around $\amg=-1.6$. For smaller $\amg$'s, it shrinks
very quickly and disappears just after $\amg=-1.62$. The fact that
the allowed regions are narrow strips along the lines with
$\Dst=0$ indicates that the main mechanism which reduces $\Omx$
below $0.125$ is the $\xstau$ coannihilations. Namely, the
dominant processes are the $\tilde\tau_2\tilde\tau_2^*$
coannihilations to $b\bar{b}$ and $\tau\bar\tau$ which are
enhanced by the $s$-channel exchange of $H$, with
$\Delta_H\simeq1.1$ -- see also \Tref{tab2}.


Extending our analysis to various $\tnb$'s for $\Dst=0$ we can
obtain a more spherical view of the overall allowed region of the
model for $\mgo$. This is because $\Dst=0$ ensures the maximal
possible reduction of $\Omx$ due to the $\xstau$ coannihilation
and so we find the maximal $\mg$ or $\mx$ allowed by \Eref{omgb}
for a given value of $\amg$. The relevant allowed (hatched)
regions in the $\mg-\amg$ plane are displayed in the right plot of
Fig.~\ref{fig:tanb48m12m0}. The right boundaries of the allowed
regions correspond to $\Omx=0.125$, while the left ones saturate
the bound on $\bsmm$ in Eq.~(\ref{bmmb}). The almost horizontal
upper boundaries correspond to the sudden shrinking of the allowed
areas which is due to the weakening of the $H$-pole effect as
$A_0/M_{1/2}$ drops below a certain value for each $\tan\beta$.
The lower left boundary of the areas for $\tan\beta=44$, 45, and
46 comes for the lower bound on $m_h$ in Eq.~(\ref{mhb}), while
the somewhat curved, almost horizontal, part of the lower boundary
of the area for $\tan\beta=44$ originates from Eq.~(\ref{omgb}).
The dot-dashed and dashed lines correspond to $m_h=125$ and
$126~{\rm GeV}$ respectively. We see that the $m_h$'s which are
favored by LHC can be readily obtained for
$47\lesssim\tan\beta\lesssim50$. In the overall allowed region we
obtain $122\lesssim\mh/{\rm GeV}\lesssim 127.23$ and
$746.5\lesssim\mx/{\rm GeV}\lesssim1433$. Also $\delta
a_\mu\simeq\lf0.35-2.76 \rg\times10^{-10}$ and so, \Eref{damub} is
satisfied only at the level of 2.55 to $2.82-\sigma$.

\subsection{$\mog$ Region}
\label{res3}

\begin{figure}[!t]\vspace*{-.12in}
\hspace*{-.19in}
\begin{minipage}{8in}
\epsfig{file=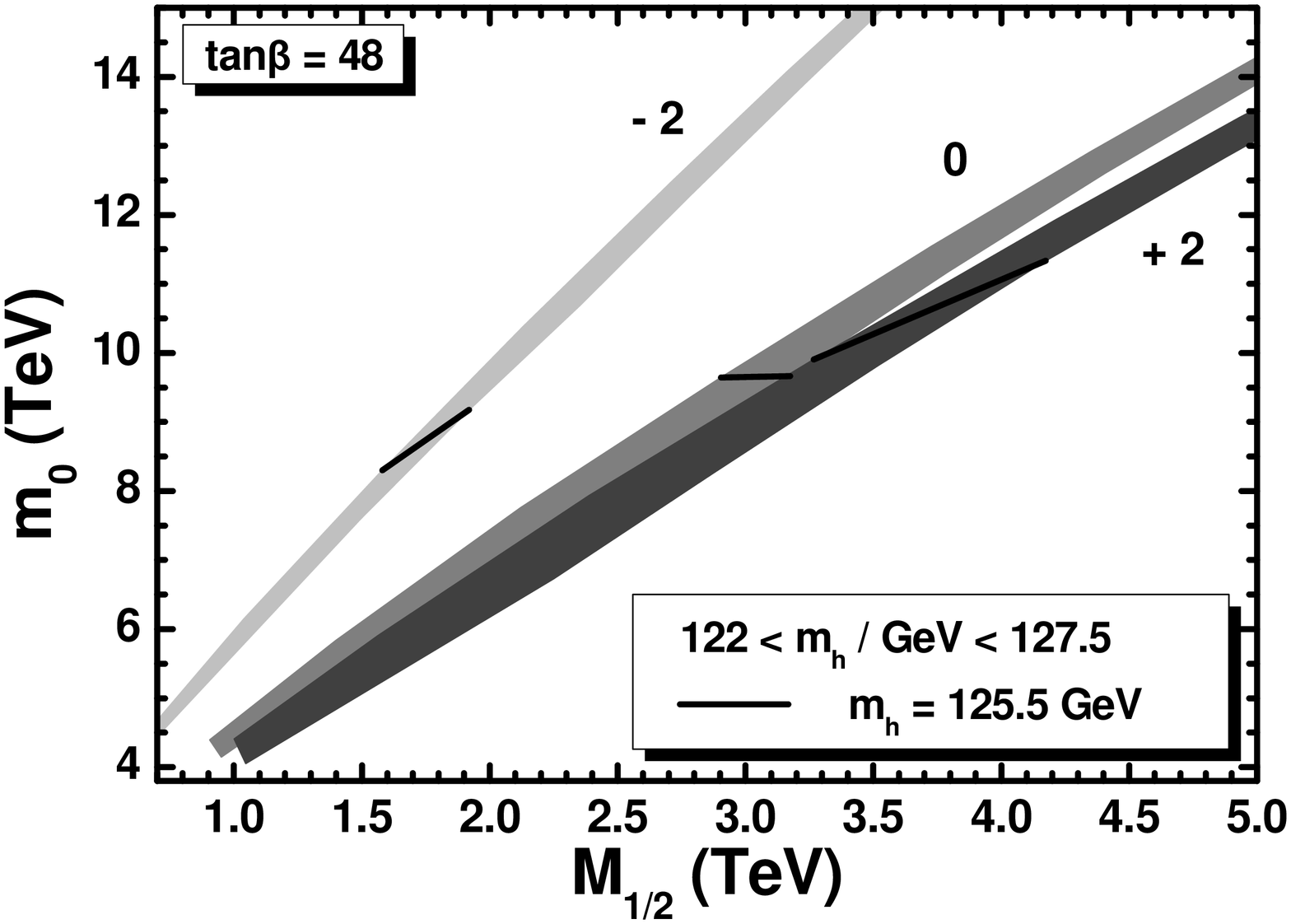,height=3.6in,angle=-90}
\hspace*{-1.2cm}
\epsfig{file=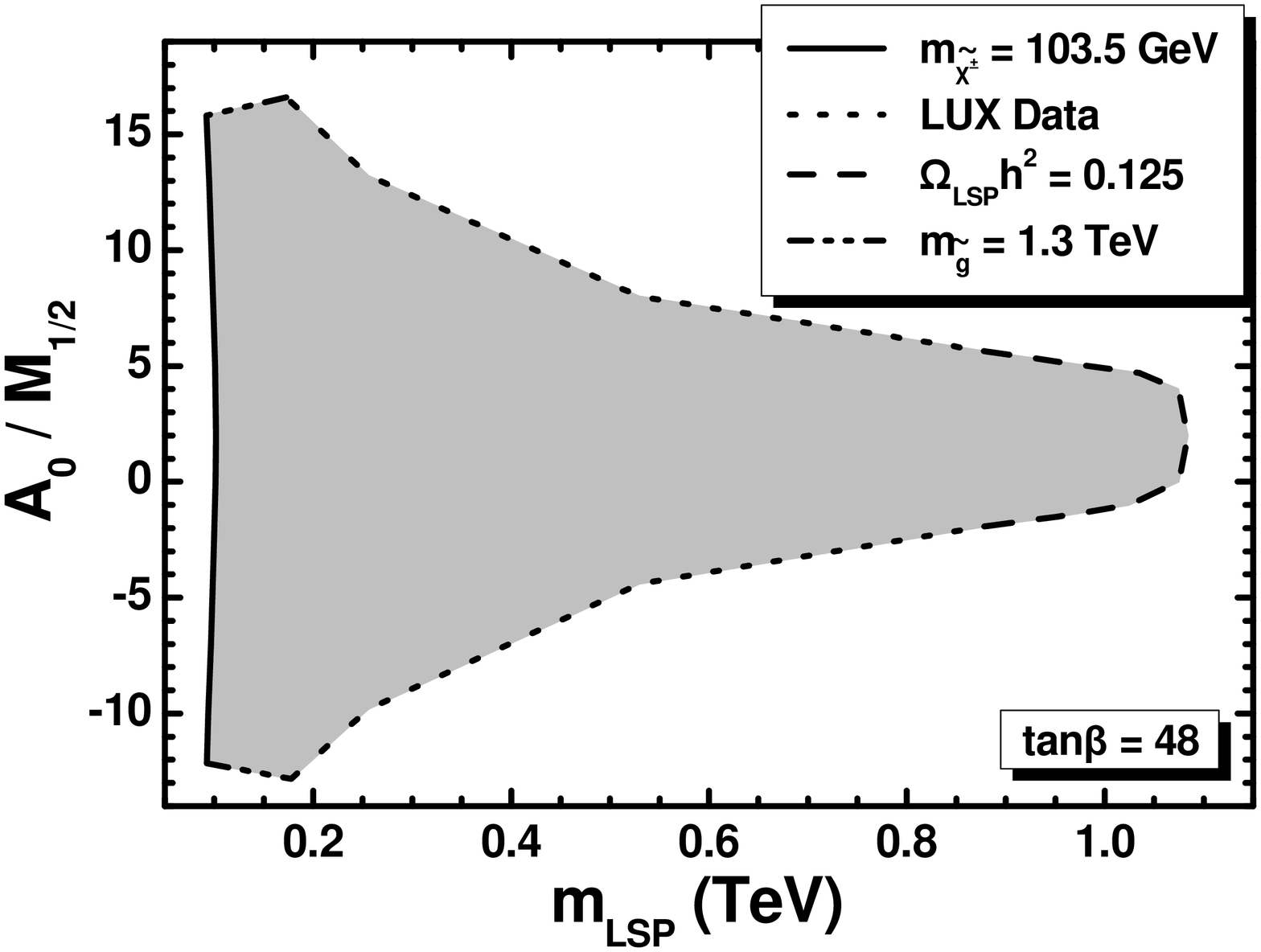,height=3.6in,angle=-90} \hfill
\end{minipage}
\hfill \caption{\sl\small Allowed (shaded) areas in the $\mg-m_0$
[$\mx-\amg$] plane (left [right] plot) for $\tan\beta=48$ and
$\mog$. In the left plot we take various $\amg$'s indicated in the
graph and the black lines correspond to $\mh=125.5\gev$. In the
right plot we set $\mh=125.5\gev$.}\label{mxA}
\end{figure}

The interplay of the various requirements of \Sref{sec:pheno} in
the $\mog$ region can be easily understood from \Fref{mxA}, where
we present the (shaded) strips in the $M_{1/2}-m_0$ plane allowed
by Eqs.~(\ref{mhb}) -- (\ref{omgb}) for $\tan\beta=48$ and several
$\amg$'s indicated in the graph -- note that no restrictions from
LUX data are applied to this plot. The upper [lower] boundary
along each of these allowed strips arises from the limit on $\mch$
[$\omg$] in Eq.~(\ref{mchb}{\ftn\sf a}) [Eq.~(\ref{omgb})]. On the
other hand, the lower limit on $\mh$ in \Eref{mhb} causes the
termination of the strips at low values of $m_0$ and $M_{1/2}$,
whereas their termination at high values of $m_0$ is put in by
hand in order to avoid shifting the SUSY masses to very large
values. The solid lines indicate solutions with $\mh=125.5\gev$ --
see \Eref{mhb}. From this figure, we easily see the main features
of the $\mog$ region: $\mo$ spans a huge range $(4-15)\tev$,
whereas $\mg$ [$\mu$] remains relatively low $(1-6)\tev$
[$(0.1-1)\tev$]. We observe also that as $\amg$ increases from
$-2$ to $2$ the allowed strip moves to larger $M_{1/2}$'s and
becomes less steep.

Varying continuously $\amg$ for $\tnb=48$, $\mh=125.5\gev$ and
taking into account the LUX data, we depict in the right plot of
\Fref{mxA} the overall allowed region of the model for $\mog$ in
the $\mx-\amg$ plane. On the solid and dashed line, the bounds on
$\mch$ in \sEref{mchb}{a} and on $\omg$ in \Eref{omgb} are
saturated, whereas the restriction from the LUX data on $\xssi$
yields the dotted boundary line. Finally, the double-dot dashed
boundary lines from the limit on $\mgl$ in \sEref{mchb}{b} provide
the maximal and minimal $\amg$'s. Note that the allowed regions
are almost symmetric about $\amg\simeq2.5$. Also, we remark that
$\mu$ remains almost constant $\simeq100\pm20 \gev$ on the solid
lines from Eq.~(\ref{mchb}), while it reaches about $1\tev$ when
the bound in Eq.~(\ref{omgb}) is saturated. Close to the latter
portion of the parameter space $\omg$ calculation is dominated by
the $\xx/\xxb-\cha$ coannihilation processes whereas in the region
where $\mch$ is near its lower limit in \sEref{mchb}{a}, the
$\xx-\xx$ annihilation processes contribute more efficiently to
the resulting  $\omg$. All in all, we obtain
$-12.8\lesssim\amg\lesssim15.8$ and
$92\lesssim{\mx/\GeV}\lesssim1084.2$. In this area  $\damu$ is
well below the lower limit in Eq.~(\ref{damub}), i.e.
$\damu\simeq(0.04-0.27)\times 10^{-10}$. Therefore, \Eref{damub}
is satisfied only at the level of $2.83$ to $2.86-\sigma$.

In the $\mog$ region, where $\xx$ has a significant higgsino
component -- see \Tref{tab2} --, $\ssi$ is dominated by the
$t$-channel Higgs-boson-exchange diagram contributing to the
neutralino-quark elastic scattering process -- for the relevant
tree-level interaction terms see e.g. the appendix of \cref{nova}.
Especially for large $\tan\beta$'s, which is the case here, the
couplings of $H$ to down-type quarks are proportional to
$\tan\beta$ and so are the dominant ones. More explicitly, $\ssi$
behaves as
\begin{equation}
 \label{ssiN} \ssi\propto \tan^2\beta|N_{1,1}|^2|N_{1,3}|^2/m_H^4,
\end{equation}
where $N_{1,1}$, $N_{1,2}$, and $N_{1,(3,4)}$ are the elements of
the matrix $N$ which diagonalizes the neutralino mass matrix and
express the bino, wino, and higgsino component of $\xx$,
respectively. As a consequence, $\ssi$ can be rather enhanced
compared to its value in the $\mgo$ region, where $\xx$ is a pure
bino.

\begin{figure}[!t]\vspace*{-.12in}
\hspace*{-.19in}
\begin{minipage}{8in}
\epsfig{file=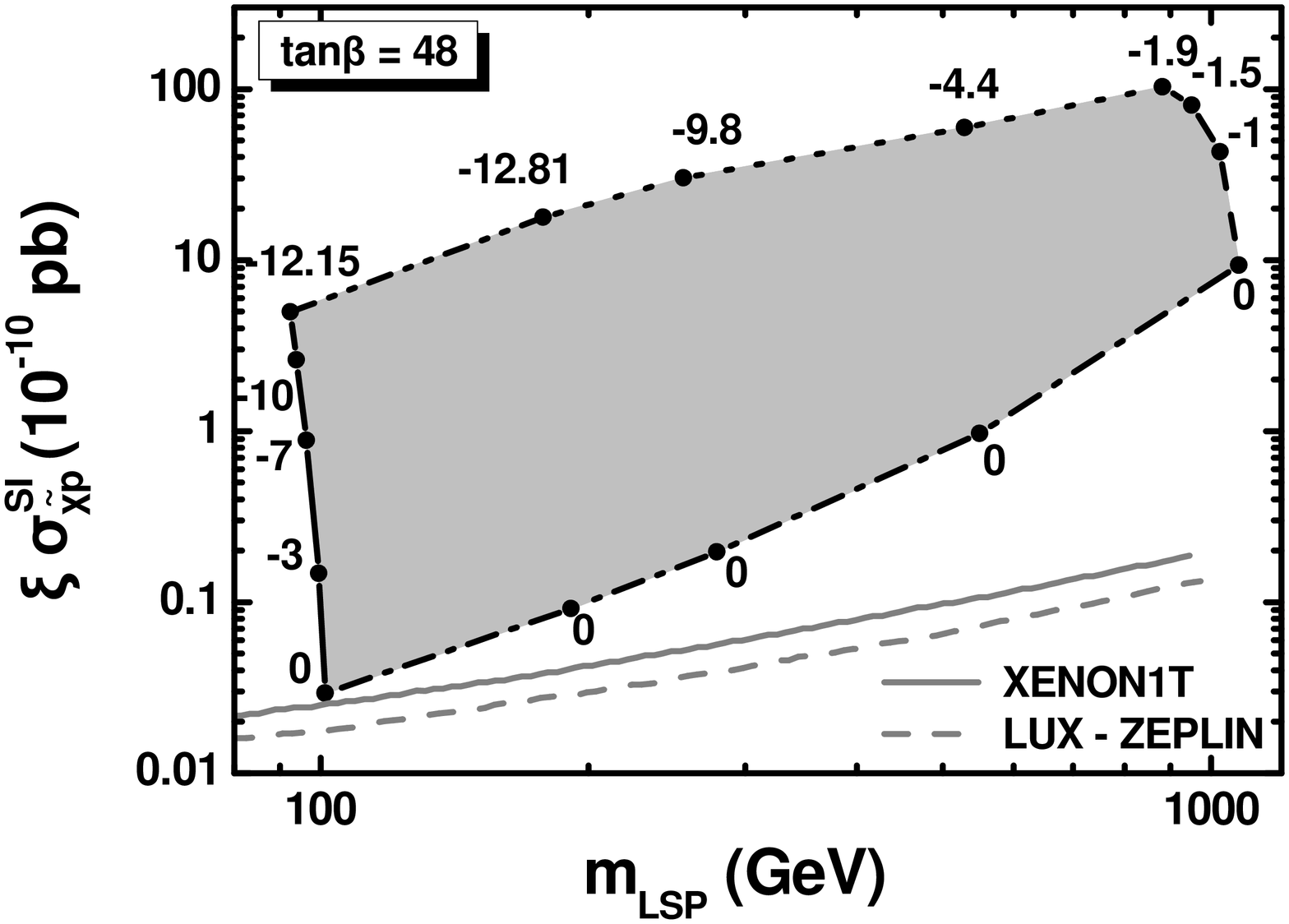,height=3.6in,angle=-90}
\hspace*{-1.2cm}
\epsfig{file=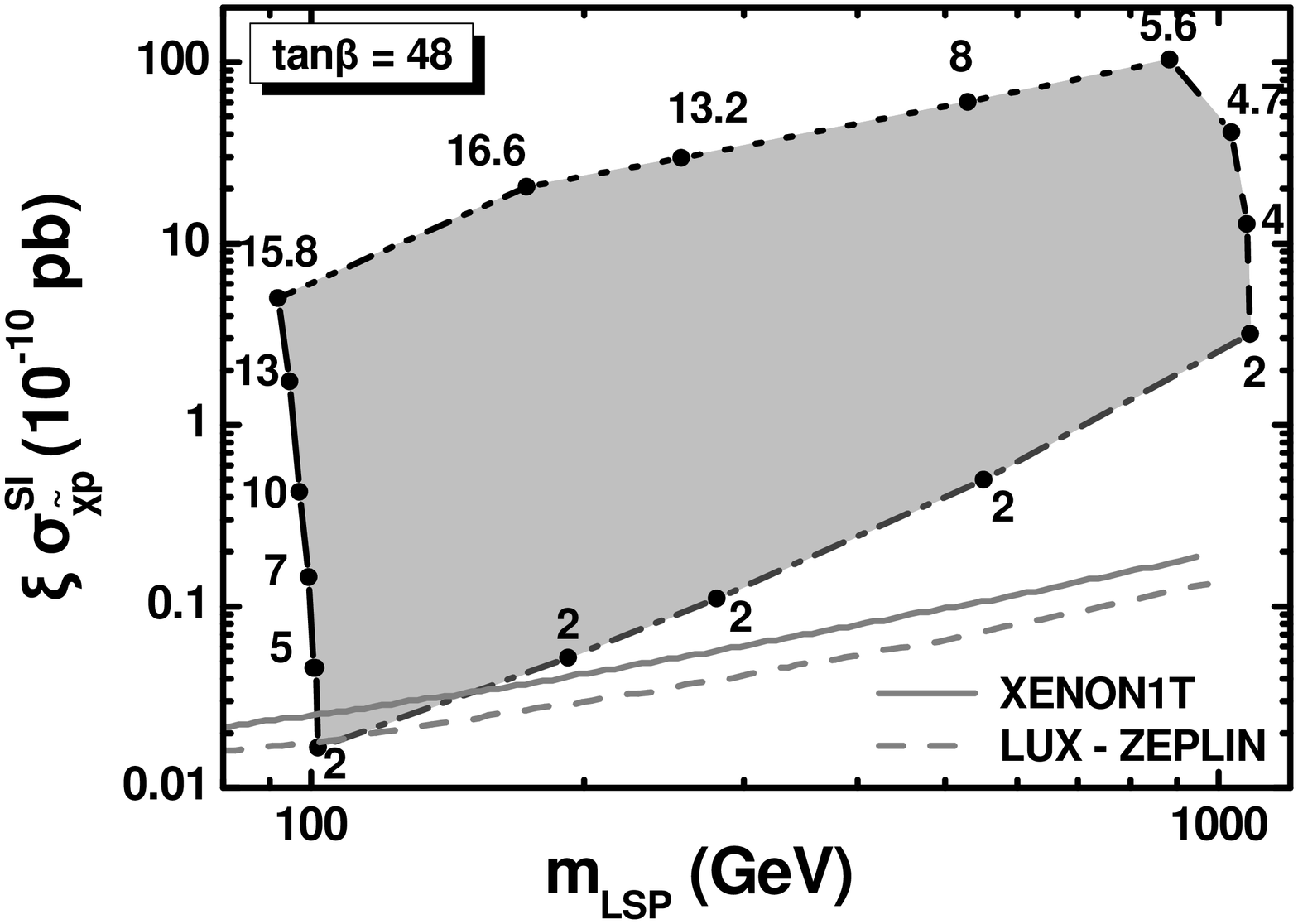,height=3.6in,angle=-90} \hfill
\end{minipage}\vspace*{-.1in}
\begin{center}
\epsfig{file=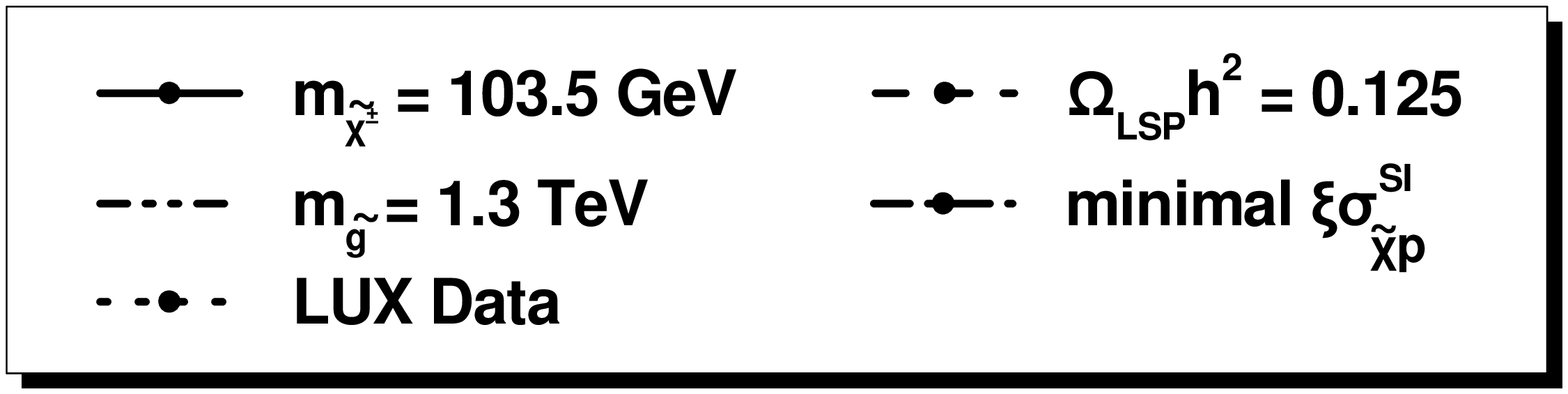,height=2.8in,angle=-90}
\end{center}
\caption{\sl\small Allowed (shaded) regions in the $\mx-\xssi$
plane for $\mog$, $m_h=125.5\gev$ and $\tan\beta=48$. The left
[right] panel corresponds to $\amg\leq 0$ [$\amg\geq 0$] and the
values of $\amg$ at the various points of the boundary lines are
indicated.  The conventions adopted for the various lines are also
shown.} \label{fig:xisigma}
\end{figure}

This conclusion can be clearly induced by Fig.~\ref{fig:xisigma},
where we show the allowed (shaded) regions in the $\mx-\xssi$
plane for $\tnb=48$. The left [right] panel corresponds to
$\amg\leq0$ [$\amg\geq0$]. The numbers on the various points of
each boundary line indicate the corresponding values of $\amg$.
The (black) solid, double dot-dashed, and dashed lines correspond
to the bounds from Eqs.~(\ref{mchb}{\sf\ftn a}),
(\ref{mchb}{\sf\ftn b}), and (\ref{omgb}), respectively -- cf.
right plot in \Fref{mxA}. The dotted lines arise from the LUX
data, whereas the dot-dashed lines give the lowest possible
$\xssi$ in each case, with $\xi$ ranging from from about $0.013$
to $1$ (along the dashed line). From these graphs, we infer that,
the maximal [minimal] $\xssi$ is located in the upper right
[lowest left] corner of the allowed regions,  at the junction
point of the dashed and dotted [dot-dashed and solid] lines. The
overall minimum of $\xssi$ is obtained in the right plot of
\Fref{fig:xisigma} whereas the maximal one is practically the same
in both plots. Namely we obtain
\beq 1.66\times10^{-12}\lesssim\xssi/{\rm pb}\lesssim
1.03\times10^{-8}.\eeq
The obtained values of $\xssi$ are within the reach of the
forthcoming experiments like XENON1T \cite{xenon1t} and LUX-ZEPLIN
\cite{supercdms}, whose the planned sensitivities are depicted in
Fig.~\ref{fig:xisigma} by a gray solid and dashed line
respectively. Note, finally, that similar results, as regards the
allowed ranges of $\mx$ and $\amg$ and the predicted $\xssi$ can
be obtained \cite{nikos15} for other $\tnb$'s in the range $40-50$
too.

\subsection{$\mgo$ Versus $\mog$ Region}
\label{res4}

Comparing the right panels of Figs.~\ref{fig:tanb48m12m0} and
\ref{mxA} we notice that the allowed areas in the $\mgo$ and
$\mog$ regions share common $\amg$'s for $\tnb=48$. Focusing on a
such $\amg$ value, e.g. $\amg=-1.5$, and selecting the remaining
input parameters so that we achieve the central values $\Omx=0.12$
and $\mh=125.5\gev$, we can implement a direct comparison between
the $\mgo$ and $\mog$ solutions. This is done in \Tref{tab2},
where we arrange the values of the input and some output
parameters, the mass spectra and some low energy observables for
two characteristic points of the allowed parameter space our
model. The various masses of the SUSY particles
(gauginos/higgsinos $\neutralino$, $\neutralino_2$,
$\neutralino_3$, $\neutralino_4$, $\neutralino_1^\pm$,
$\neutralino_2^\pm$, $\tilde{g}$, squarks $\tilde{t}_1$,
$\tilde{t}_2$, $\tilde{b}_1$, $\tilde{b}_2$, $\tilde{u}_L$,
$\tilde{u}_R$, $\tilde{d}_L$, $\tilde{d}_R$, and sleptons
$\tilde{\tau}_1$, $\tilde{\tau}_2$, $\tilde{\nu}_\tau$,
$\tilde{e}_L$, $\tilde{e}_R$) and the Higgs particles ($h$, $H$,
$H^\pm$, $A$) are given in $\TeV$ -- note that we consider the
first two generations of squarks and sleptons as degenerate.  From
the values of the various observable quantities, we can verify
that all the relevant constraints, but the one of \Eref{damub},
are met -- cf. \Sref{sec:pheno}. For the interpretation of our
results, mainly on $\omg$, we also list the values of the various
$\Delta_P$'s in \Eref{ds}, the bino, $|N_{1,1}|^2$, and the
higgsino, $|N_{1,3}|^2+|N_{1,4}|^2$, purity of $\xx$. We also
include an estimate for the EWSB fine-tuning parameter $\dew$ --
see \Sref{nat}.

\clearpage\renewcommand{\arraystretch}{0.98}
\begin{table}[!h]
\bec\begin{tabular}{|ccc|} \hline \multicolumn{3}{|c|}{\sc Input
Parameters}\\\hline
$\tan\beta$ & \multicolumn{2}{c|}{$48$} \\
$-\amg$ & \multicolumn{2}{c|}{$1.5$}\\
$\mg/\TeV$ & $2.821$ &$2.157$ \\
$m_0/\TeV$ &$2.522$ &$9.219$\\ \hline
\multicolumn{3}{|c|}{\sc Output Parameters}\\\hline
$h_t/h_\tau(M_{\rm GUT})$ & $1.117$ & $1.107$\\
$h_b/h_\tau(M_{\rm GUT})$ & $0.623$ & $0.763$\\
$h_t/h_b(M_{\rm GUT})$ & $1.792$ & $1.45$ \\\hline
$\mu/\TeV$ & $3.514$ & $0.936$ \\
$\dew$ &$2972$ &$216.2$\\\hline
$\Dst~(\%)$ & $0.35$ & $615$ \\
$\Delta_H~(\%)$ & $1.14$ & $94$ \\ \hline
$\dch~(\%)$&$89.96$ & $1.56$ \\
$\dnt~(\%)$&$89.95$ & $1.928$ \\\hline
$|N_{1,1}|^2$&$1$ &$0.145$\\
$|N_{1,3}|^2+|N_{1,4}|^2$&$0$ &$0.852$\\\hline
\multicolumn{3}{|c|}{\sc Masses in ${\rm TeV}$ of Sparticles and
Higgses}\\\hline
$\tilde\chi, \tilde\chi_2^{0}$& $1.305, 2.478$ &{\bf\boldmath $0.943, 0.960$} \\
$\tilde{\chi}_{3}^{0}, \tilde{\chi}_{4}^{0}$ &$3.509, 3.512$ &$1.034, 1.959$ \\
$\tilde{\chi}_{1}^{\pm}, \tilde{\chi}_{2}^{\pm}$ &$2.479, 3.512$ &{\bf\boldmath  $0.956$}, $1.959$ \\
$\tilde{g}$&$6$ &$4.936$  \\ \hline
%
$\tilde{t}_1, \tilde{t}_2$ &$3.998, 4.729$ &$6.309, 7.270$ \\
$\tilde{b}_1, \tilde{b}_2$ &$4.692, 4.772$ &$7.267, 7.887$ \\
$\tilde{u}_{L}, \tilde{u}_{R}$&$5.880, 5.625$ &$10.1, 10.004$ \\
$\tilde{d}_{L}, \tilde{d}_{R}$& $5.880, 5.592$ &$10.1, 9.992$
\\\hline
$\tilde\tau_1, \tilde\tau_2$&$1.309, 2.661$ &$6.749, 8.202$ \\
$\tilde{e}_L, \tilde{e}_R$&$3.162, 2.744$ &$9.359, 9.276$ \\
$\tilde\nu_\tau, \tilde{\nu}_{e}$ &$2.656, 3.160$ &$8.201, 9.359$ \\
\hline
%
$h, H$&$0.1255, 2.640$ &$0.1255, 3.67$ \\
$H^{\pm}, A$&$2.641, 2.640$  &$3.671, 3.67$  \\\hline
\multicolumn{3}{|c|}{\sc Low Energy Observables}\\\hline
%
$10^4\bsg$ &$3.27$  &$3.3$ \\
$10^9\bsmm$   &$3.74$ &$3.02$ \\
$\btn$   &$0.984$  &$0.991$ \\
$10^{10}\damu$  &$0.68$&$0.227$ \\\hline
$\Omx$ &\multicolumn{2}{c|}{$0.12$}\\
$\ssi /  \pb$&$3.35\cdot 10^{-12}$&$7.28\cdot 10^{-9}$ \\
$\ssd /  \pb$&$6.67\cdot10^{-10}$&$8.57\cdot 10^{-6}$ \\
\hline
\end{tabular}\eec
\caption{\sl\small Input/output parameters, sparticle and Higgs
masses, and low energy observables for $\tnb=48$, $\amg=-1.5$ and
$\mgo$ (second column) or $\mog$ (third column).} \label{tab2}
\end{table}

The results of the two columns of \Tref{tab2} reveal that the
$\mgo$ and $\mog$ solutions exhibit a number of important
differences: First of all, in the $\mog$ region, $m_0$ acquires
considerably larger values, while $\mu$ remains quite smaller than
its value for $\mgo$. As a consequence, the Higgs bosons ($H$,
$H^\pm$, and $A$) acquire larger masses and the whole sparticle
spectrum, with the exception of the neutralinos and charginos,
becomes heavier. As a by-product, the various observables besides
$\omg$ acquire values closer to non SUSY ones. E.g., $\damu$ is
even smaller than in the $\mgo$ region. Note that the latter
region is tightly constrained by $\bsmm$, which is well suppressed
for $\mog$. Similar values for the ratios $h_m/h_n$ with
$m,n=t,b,\tau$ are obtained in both cases with results slightly
closer to unity for $\mog$.

\renewcommand{\arraystretch}{1.1}
\begin{floatingtable}[l]
\begin{tabular}{|c|c|}\hline
\multicolumn{2}{|c|}{\sc $\mgo$ Region}\\ \hline
$\staua\staua^*\rightarrow b\bar b$ &$69\%$\\
$\staua\staua^*\rightarrow \tau\bar\tau $&$15\%$\\ \hline
\multicolumn{2}{|c|}{\sc $\mog$ Region}\\ \hline
$\xx\cha\rightarrow  u\bar d $&$18\%$\\
$\xxb\cha\rightarrow  u\bar d$ &$8\%$\\
$\xx\cha\rightarrow  t\bar b$ &$7\%$\\
$\xx\cha\rightarrow  \nu_e\bar e$ &$6\%$\\\hline
$\xx\xx\rightarrow  W^-W^+$ &$6\%$\\
$\xx\xx\rightarrow  ZZ $&$5\%$\\ \hline
\end{tabular}
\caption{\sl\small Processes which contribute to $1/\Omx$ more
than $5\%$ and their relative contributions for $\mgo$ and
$\mog$.} \label{tab3}
\end{floatingtable}
To shed more light on the mechanisms which ensure $\omg$
compatible with \Eref{omgb} for $\mgo$ and $\mog$, we arrange in
\Tref{tab3} the relative contributions beyond 5$\%$ to $1/\omg$ of
the various (co)annihilation processes  for the inputs of
\Tref{tab2}. From these results, we infer that a synergy between
$\xstau$ coannihilation and the $H$-funnel mechanism is well
established in the $\mgo$ region thanks to the quite suppressed
$\dstau$ and $\Delta_H$ achieved. On the other hand, for $\mog$
the $\xx/\xxb-\cha$ coannihilations are activated because of the
low $\dch$'s and $\dnt$'s. However, in the latter case the
annihilation channels to $W^-W^+$ and $ZZ$ conserve their
importance due to the large higgsino mixing of $\xx$. Due to this
fact $\mx$ is confined close to $\mu$ and, as a bonus, the
resulting $\ssi$'s are accessible to the forthcoming experiments
\cite{xenon1t,supercdms}. On the other hand, for $\mgo$, $\xx$ is
an almost pure bino with $\mx\simeq\mg/2$ and $\ssi$ well below
the sensitivity of any planned experiment. In both regions,
finally, $\ssd$ turns out to be much lower than the reach of
IceCube \cite{icecube} (assuming $\xx-\xx$ annihilation into
$W^+W^-$) and the expected limit from the large DMTPC detector
\cite{brown}. Therefore, the LSPs predicted by our model can be
detectable only by the future experiments which will release data
on $\ssi$ and for $\mog$.

\begin{figure}[!t]\vspace*{-.12in}
\hspace*{-.19in}
\begin{minipage}{8in}
\epsfig{file=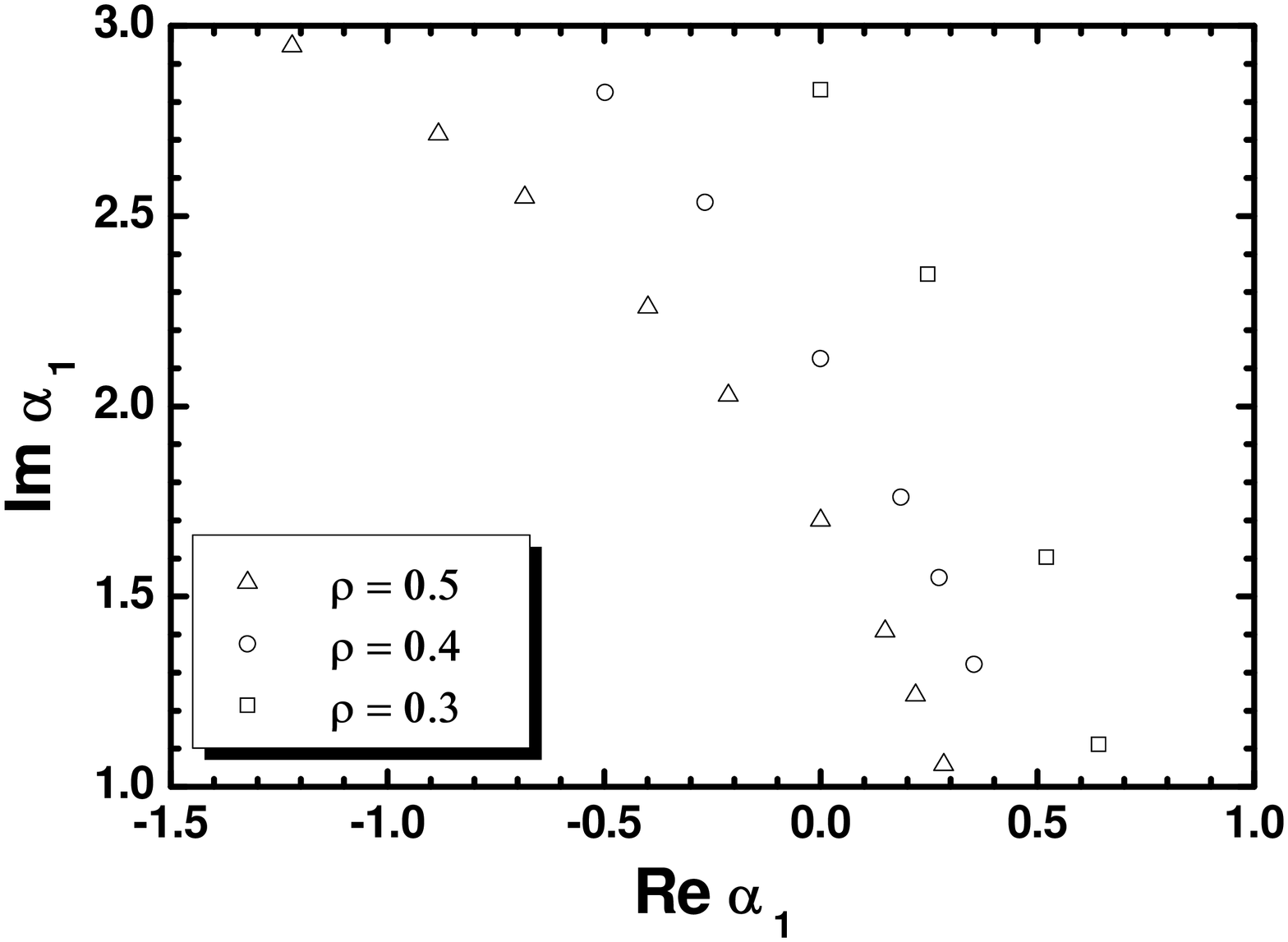,height=3.6in,angle=-90}
\hspace*{-1.2cm}
\epsfig{file=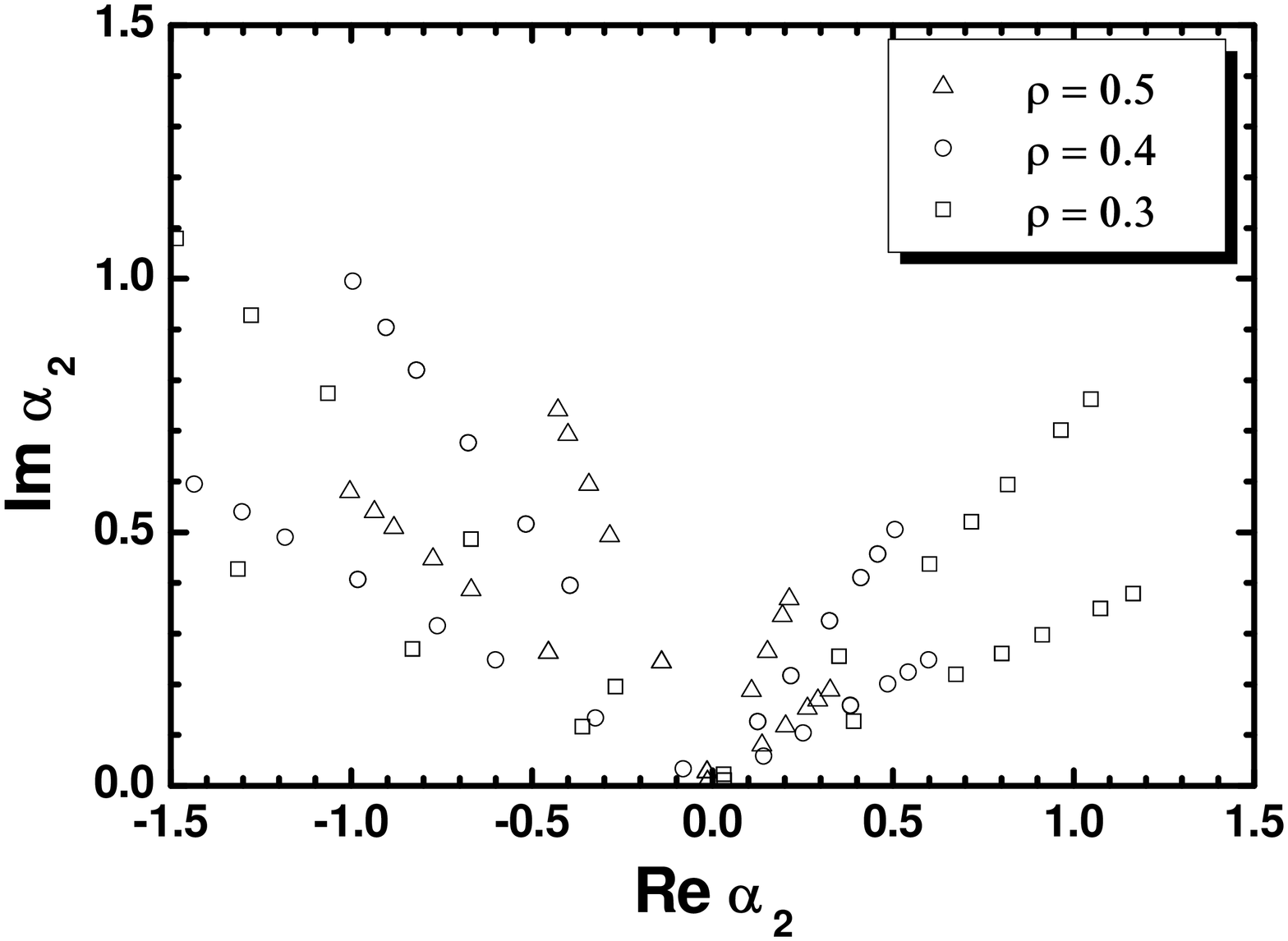,height=3.6in,angle=-90} \hfill
\end{minipage}
\hfill \hspace*{-.19in}
\begin{minipage}{8in}
\epsfig{file=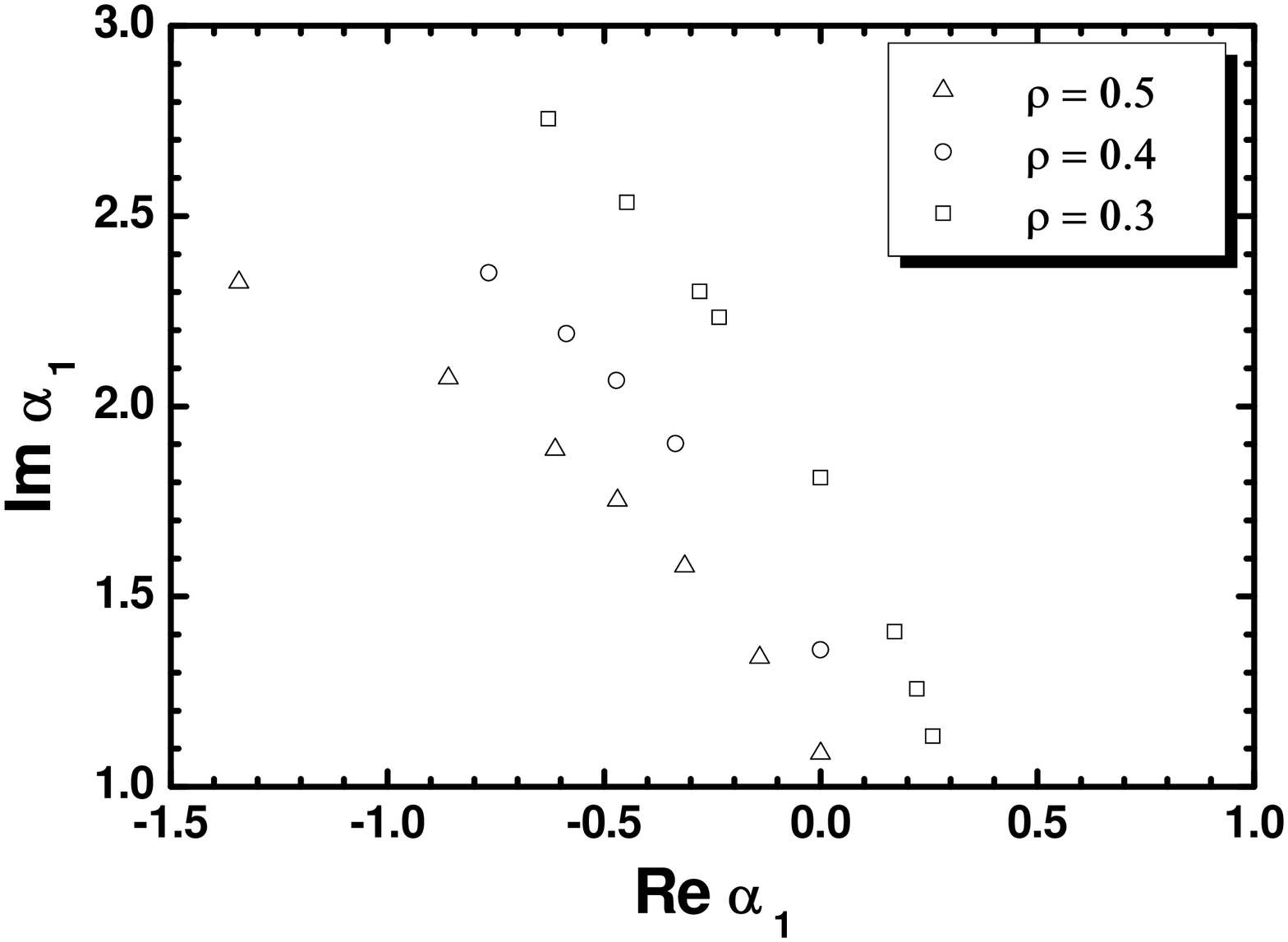,height=3.6in,angle=-90}
\hspace*{-1.2cm}
\epsfig{file=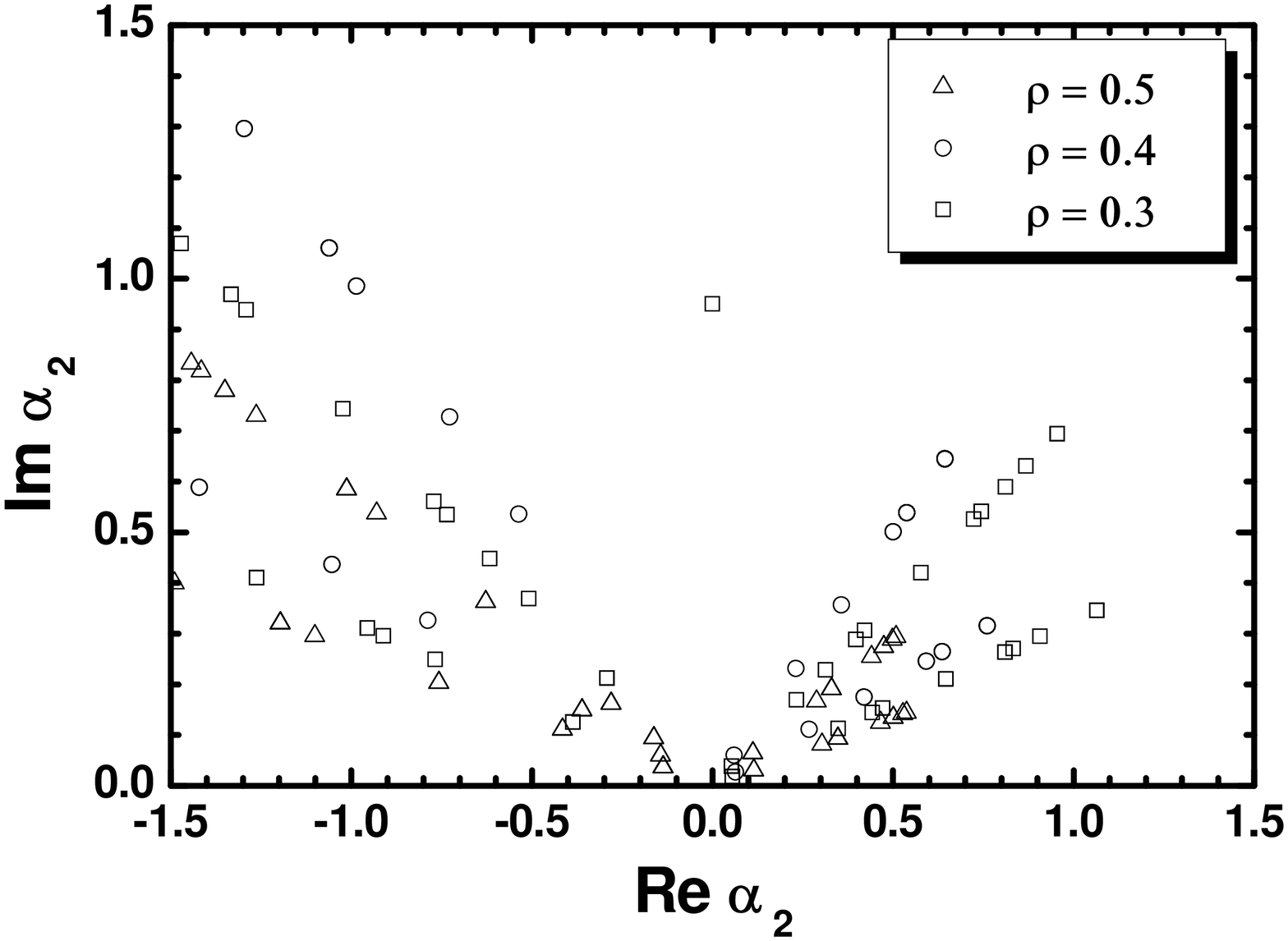,height=3.6in,angle=-90} \hfill
\end{minipage}
\caption{\sl\ftn The complex parameters $\alpha_1$ and $\alpha_2$
for various $\rho$'s indicated in the graphs for $tan\beta=48$,
$\amg=-1.5$, $m_h=125.5~{\rm GeV}$ and $\mgo$ (upper plots) and
$\mog$ (lower plots) -- see Table 2.} \label{fig:a1a2}
\end{figure}

\section{Deviation from Yukawa Unification} \label{yuk}

In the overall allowed parameter space of our model
\cite{nikos13,nikos15}, we find the following ranges for the
ratios $h_m/h_n$ with $m,n=t,b,\tau$:
\beqs \bea && 0.98\lesssim{h_t}/{h_\tau}\lesssim1.29,~0.60\lesssim
{h_b}/{h_\tau}\lesssim0.65,~\text{and}~
1.62\lesssim{h_t}/{h_b}\lesssim2~~~\mbox{for}~~~\mgo;\label{specratio1}\\
&& 1.00\lesssim{h_t}/{h_\tau}\lesssim1.50,~0.75\lesssim
{h_b}/{h_\tau}\lesssim0.79,~\text{and}~
1.20\lesssim{h_t}/{h_b}\lesssim2~~~\mbox{for}~~~\mog
\label{specratio2}\eea\eeqs
We observe that, the required deviation from YU is not so small
and turns out to be comparable to the one obtained in the
monoparametric case -- cf. \cref{nikos11}. In spite of this, the
restrictions from YU are not completely lost but only somewhat
weakened. Actually, our model is much closer to YU than generic
models with lower $\tan\beta$'s where the Yukawa coupling
constants can differ even by orders of magnitude. Also, the
deviation from YU is generated by Eq.~(\ref{yquc}) in a natural,
systematic, controlled and well-motivated way.

To show explicitly it, we below extract values of $\rho$,
$\alpha_1$, and $\alpha_2$, which leads to the ratios $h_m/h_n$
with $m,n=t,b,\tau$, encountered in the characteristic examples
presented in Table~\ref{tab2}. Since from Eq.~(\ref{yquc}) we have
only two equations and five real unknowns we can find infinitely
many solutions. Some of these solutions are shown
Fig.~\ref{fig:a1a2} for $\mgo$ (upper plots) and $\mog$ (lower
plots) and various $\rho$'s indicated therein. Since the equation
for $h_b/h_\tau$ depends only on the combination $\rho\alpha_1$
its solutions, for fixed $\rho$, lie on a certain curve in the
$\alpha_1$ complex plane, as shown in the left upper and lower
panels of Fig.~\ref{fig:a1a2}. For each $\alpha_1$ and $\rho$ in
these panels, we find various $\alpha_2$'s, depicted in the right
panels of Fig.~\ref{fig:a1a2}, solving the equation for
$h_t/h_\tau$. Observe that the equation for $h_t/h_\tau$ depends
separately on $\alpha_2$ and $\rho$ and, thus, its solutions do
not follow any specific pattern in the $\alpha_2$ complex plane.
Scanning the range of $\rho$ from $0.3$ to $3$ and we can find
solutions in the $\alpha_1$ and $\alpha_2$ planes only for the
lower values of this parameter (up to about $0.6$) for both
allowed regions of the model. These solutions are very similar to
the ones displayed in Fig.~\ref{fig:a1a2} for all the possible
values of the ratios of $h_m/h_n$ with $m,n=t,b,\tau$ allowed by
the constraints of \Sref{sec:pheno}. Consequently, we can safely
conclude that these ratios can be readily obtained by a multitude
of natural choices of the parameters $\rho$, $\alpha_1$, and
$\alpha_2$ everywhere in the overall allowed parameter space of
the model.

\section{Naturalness of the EWSB}\label{nat}

The fact that, in our model, $\mg$, $m_0$, and $\mu$ generally
turn out to be of the order of a few $\TeV$ puts under some stress
the naturalness of the radiative EWSB giving rise to the so-called
little hierarchy problem. To quantified somehow this issue we
introduce the EWSB fine-tuning parameter
\begin{equation}
\label{dew}
\dew\equiv\mathrm{max}\left(\frac{|C_i|}{M_Z^2/2}\right)
~~~\mbox{with}~~~ \lf C_{\mu}, C_{H_1} C_{H_2}\rg= \lf-\mu^2,
\frac{m_{H_1}^2}{\tan^2\beta-1},
-\frac{m_{H_2}^2\tan^2\beta}{\tan^2\beta-1}\rg. \eeq
Here $i=\mu,H_1,H_2$ and $m_{H_j}$ is the soft SUSY breaking mass
of $H_j$ with $j=1,2$. In most of the parameter space explored,
$\dew$ is dominated by the term $C_\mu$.

\begin{figure*}[t!]
\centerline{\epsfig{file=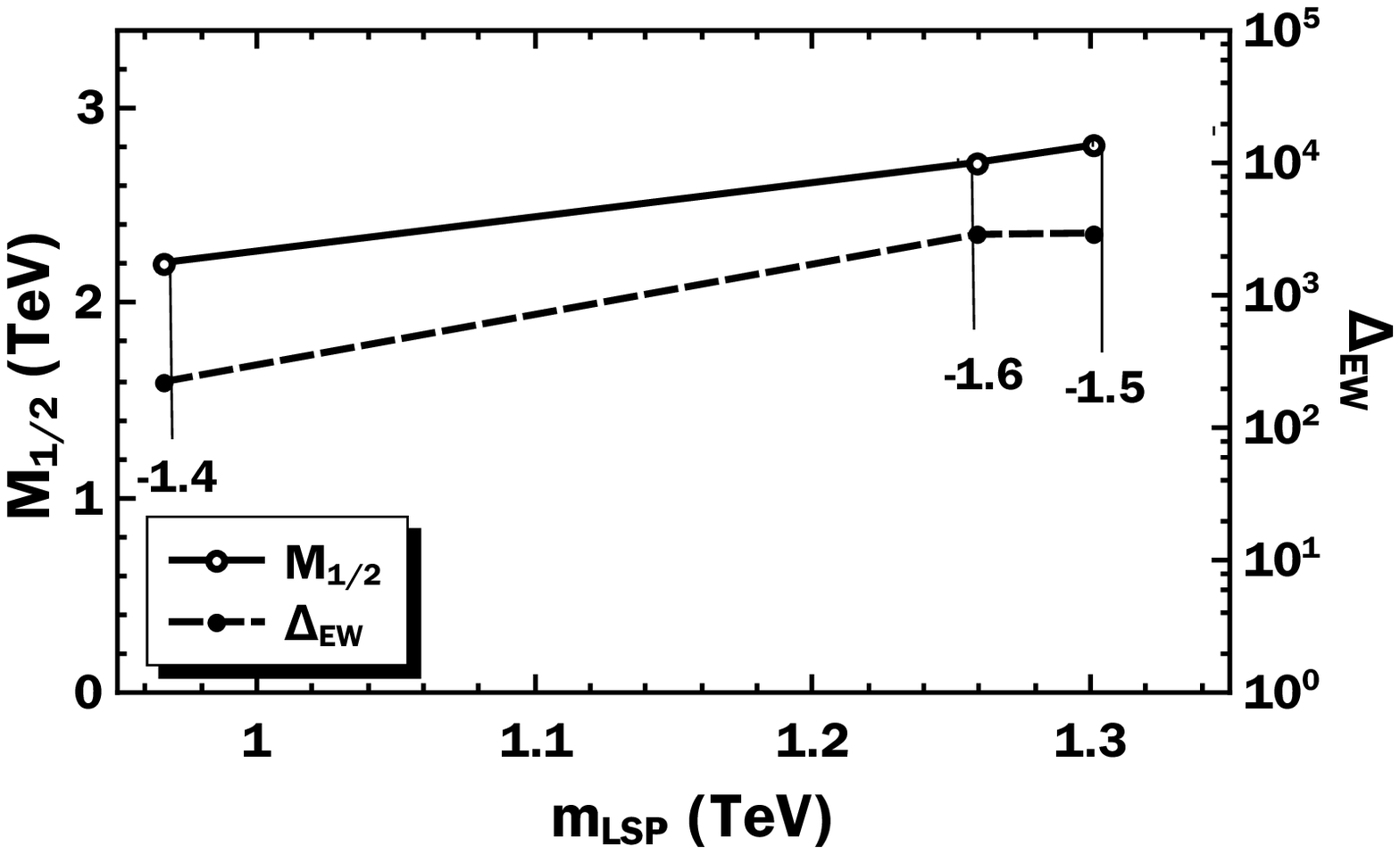,height=2.1in}
\epsfig{file=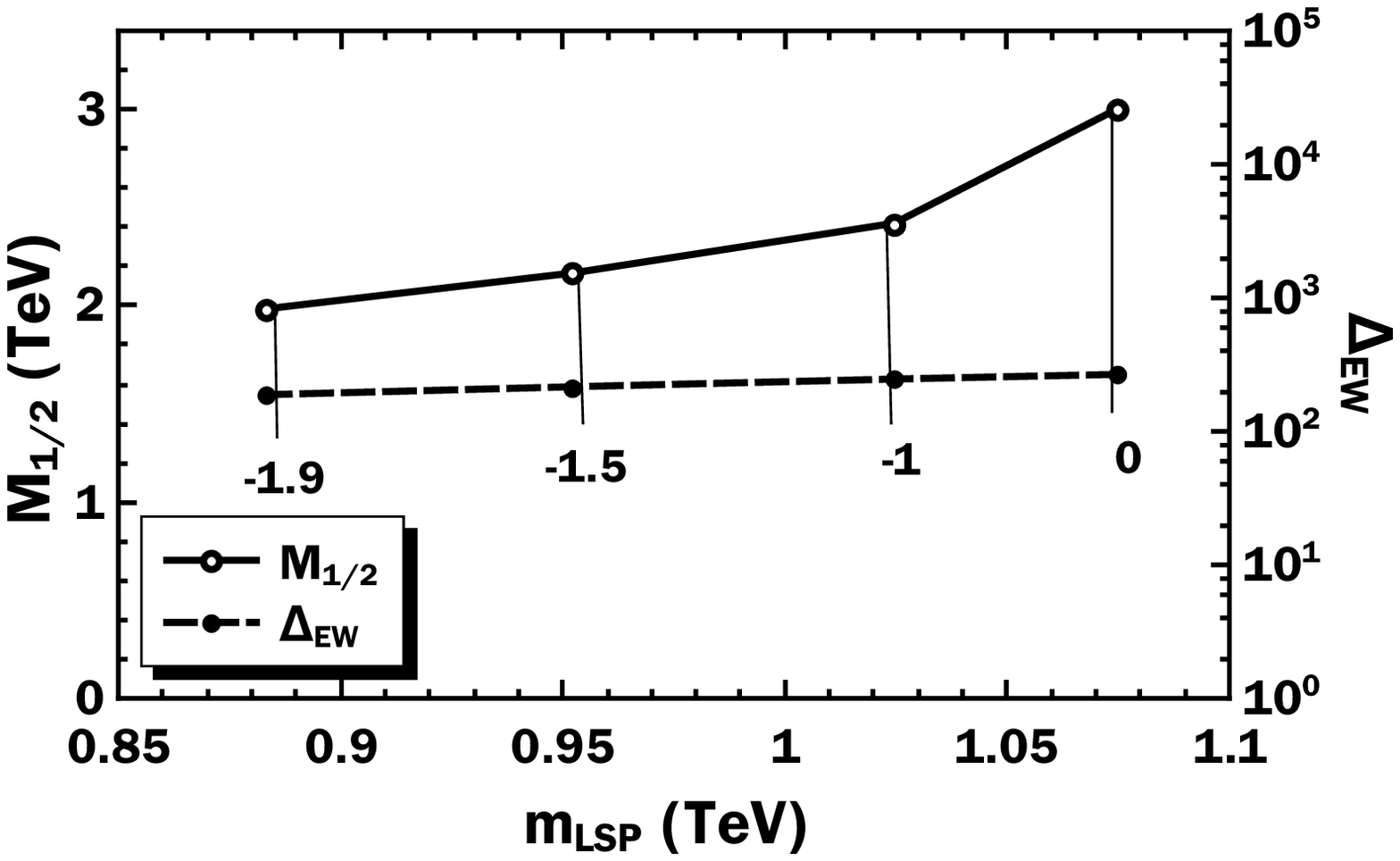,height=2.1in}}
\caption{\ftn\sl $M_{1/2}$ and $\dew$ as functions of $\mx$ for
$\mh=125.5~\GeV$, $\omg=0.125$ and various $\amg$'s indicated on
the curves for the $\mgo$ (left panel) and the $\mog$ (right
panel) region of the model.} \label{fig:finetuning}
\end{figure*}

Focusing on the values of the parameters which ensure
$\omg\simeq0.125$, we present, in the left [right] panel of
\Fref{fig:finetuning}, $\mg$ (solid line) and $\dew$ (dashed line)
as functions of $\mx$ for $\mgo$ [$\mog$], $\tnb=48$,
$\mh=125.5~\GeV$, and negative $\amg$'s indicated in the graphs.
The $\amg$'s fulfilling these conditions are shown in Fig.~3 of
\cref{nikos13} for $\mgo$ and in the left panel of
\Fref{fig:xisigma} for $\mog$. We clearly see that, in both
regions, the resulting $\dew$ is almost constant and $\dew\sim
2000$ for $\mgo$ whereas $\dew\sim 200$ for $\mog$ in agreement
with the values shown in \Tref{tab2}. In other words, $\dew$ in
the $\mgo$ area becomes about a factor of ten larger than its
value in the $\mog$ area despite the fact that the resulting
$\mx$'s are comparable. The crucial difference between the two
regions, though, is the lower $\mu$'s encountered for $\mog$ --
see \Tref{tab2} -- which largely influences the $\dew$ computation
-- see \Eref{dew}. We can conclude, therefore, that the $\mog$
solutions are more natural regarding the EWSB fine-tuning than
those for $\mgo$.

\section{Conclusions}\label{con}

We investigated the compatibility of the generalized asymptotic
YQUCs in \Eref{yquc}, which yield acceptable masses for the
fermions of the third family, with the CMSSM for $\mu>0$ and
$40\leq\tnb\leq50$. We imposed phenomenological constraints
originating from the mass of the lightest neutral CP-even Higgs
boson, the lower bounds on the masses of the sparticles, and
$B$-physics. We also considered cosmological constraints coming
from $\Omx$ and the LUX data on $\xssi$.

We found that $\xx$ can act as a CDM candidate in the following
two separated regions classified in the HB of the EWSB:

\begin{itemize}

\item The $\mgo$ region, where the LSP turns out to be an
essentially pure bino and $\Omx$ is reduced efficiently, thanks to
$H$-pole enhanced stau-antistau coannihilations, so that it is
compatible with the recent data on $\bsmm$. The LHC preferred
values $m_h\simeq(125-126)~{\rm GeV}$ can be accommodated for
$48\lesssim\tnb\lesssim50$, whereas $\mx$ comes out to be large
($\sim 1~{\rm TeV}$). As a consequence, the $\xx$ direct
detectability is very difficult and the EWSB fine-tuning becomes
rather aggressive since $\dew^{-1}\sim0.035\%$.

\item The $\mog$ region, where the LSP is a bino-higgsino
admixture and has an acceptable $\Omx$ thanks to $\xx-\xx$
annihilations (for low $\mx$'s) and the $\xx/\xxb-\cha$
coannihilations (for large $\mx$'s). Fixing $\mh=125.5\gev$
favored by the LHC, we found a wider allowed parameter space with
$40\lesssim\tnb\lesssim50$, $-11\lesssim\amg\lesssim15$,
$0.09\lesssim\mx/\TeV\lesssim1.1$ and milder EWSB fine-tuning
since $\dew^{-1}\sim0.5\%$. The LSP is possibly detectable in the
planned CDM direct search experiments which look for $\ssi$.

\end{itemize}

In both cases above, the restriction on $\damu$ is only satisfied
at a level of above $2-\sigma$ and the required deviation from YU
can be easily attributed to a multitude of natural values of the
relevant parameters within a PS SUSY GUT model.

\acknowledgments This research was supported from the MEC and
FEDER (EC) grants FPA2011-23596 and the Generalitat Valenciana
under grant PROMETEOII/2013/017.

\def\ijmp#1#2#3{{\emph{Int. Jour. Mod. Phys.}}
{\bf #1},~#3~(#2)}
\def\plb#1#2#3{{{Phys. Lett.  B }}{\bf #1},~#3~(#2)}
\def\zpc#1#2#3{{Z. Phys. C }{\bf #1},~#3~(#2)}
\def\prl#1#2#3{{\emph{Phys. Rev. Lett.} }
{\bf #1},~#3~(#2)}
\def\rmp#1#2#3{{Rev. Mod. Phys.}
{\bf #1},~#3~(#2)}
\def\prep#1#2#3{{Phys. Rep. }{\bf #1},~#3~(#2)}
\def\prd#1#2#3{{{Phys. Rev.  D} }{\bf #1},~#3~(#2)}
\def\npb#1#2#3{{{Nucl. Phys.} }{\bf B#1},~#3~(#2)}
\def\npps#1#2#3{{Nucl. Phys. B (Proc. Sup.)}
{\bf #1},~#3~(#2)}
\def\mpl#1#2#3{{Mod. Phys. Lett.}
{\bf #1},~#3~(#2)}
\def\arnps#1#2#3{{Annu. Rev. Nucl. Part. Sci.}
{\bf #1},~#3~(#2)}
\def\sjnp#1#2#3{{Sov. J. Nucl. Phys.}
{\bf #1},~#3~(#2)}
\def\jetp#1#2#3{{JETP Lett. }{\bf #1},~#3~(#2)}
\def\app#1#2#3{{Acta Phys. Polon.}
{\bf #1},~#3~(#2)}
\def\rnc#1#2#3{{Riv. Nuovo Cim.}
{\bf #1},~#3~(#2)}
\def\ap#1#2#3{{Ann. Phys. }{\bf #1},~#3~(#2)}
\def\ptp#1#2#3{{Prog. Theor. Phys.}
{\bf #1},~#3~(#2)}
\def\apjl#1#2#3{{Astrophys. J. Lett.}
{\bf #1},~#3~(#2)}
\def\n#1#2#3{{Nature }{\bf #1},~#3~(#2)}
\def\apj#1#2#3{{Astrophys. J.}
{\bf #1},~#3~(#2)}
\def\anj#1#2#3{{Astron. J. }{\bf #1},~#3~(#2)}
\def\mnras#1#2#3{{MNRAS }{\bf #1},~#3~(#2)}
\def\grg#1#2#3{{Gen. Rel. Grav.}
{\bf #1},~#3~(#2)}
\def\s#1#2#3{{Science }{\bf #1},~#3~(#2)}
\def\baas#1#2#3{{Bull. Am. Astron. Soc.}
{\bf #1},~#3~(#2)}
\def\ibid#1#2#3{{\it ibid. }{\bf #1},~#3~(#2)}
\def\cpc#1#2#3{{Comput. Phys. Commun.}
{\bf #1},~#3~(#2)}
\def\astp#1#2#3{{Astropart. Phys.}
{\bf #1},~#3~(#2)}
\def\epjc#1#2#3{{Eur. Phys. J. C}
{\bf #1},~#3~(#2)}
\def\nima#1#2#3{{Nucl. Instrum. Meth. A}
{\bf #1},~#3~(#2)}
\def\jhep#1#2#3{{\emph{JHEP} }
{\bf #1},~#3~(#2)}
\def\jcap#1#2#3{{\emph{JCAP} }
{\bf #1},~#3~(#2)}

\newcommand{\hepth}[1]{{\tt hep-th/#1}}
\newcommand{\hepph}[1]{{\tt hep-ph/#1}}
\newcommand{\hepex}[1]{{\tt hep-ex/#1}}
\newcommand{\astroph}[1]{{\tt astro-ph/#1}}
\newcommand{\arxiv}[1]{{\tt arXiv:#1}}
\newcommand{\etal}{{\it et al.\/}}

\end{document}